\documentclass[pdftex,twocolumn,epjc3]{svjour3}

\usepackage[colorlinks,citecolor=blue,urlcolor=blue,linkcolor=blue]{hyperref}

\usepackage[dvipsnames]{xcolor}

\usepackage[numbers,sort&compress]{natbib}

\usepackage{mathtools}

\usepackage{braket}
\usepackage{textcomp,gensymb}
\usepackage{dsfont}

\usepackage{graphicx}
\usepackage[export]{adjustbox}
\usepackage[margin=2cm]{geometry}

\makeatletter
\def\cl@chapter{}
\makeatother

\usepackage{hyperref}
\usepackage[capitalise]{cleveref}
\usepackage{csquotes}

\usepackage{tikz}
\usepackage{pgfplots}

\usepackage{algorithm}
\usepackage{algpseudocode}
\usepackage{bbold}
\usepackage[mode=buildnew]{standalone}

\usepackage{pifont}
\usepackage{verbatim}

\usepackage{amsmath}
\usepackage{epsfig}
\usepackage{color}
\usepackage{graphicx}

\usepackage{xcolor}

\definecolor{p-channel}{RGB}{0,153,230}
\definecolor{d-channel}{RGB}{40,37,110}
\definecolor{c-channel}{RGB}{242,56,20}

\definecolor{vertex_fill}{RGB}{125,125,125}

\usepackage{tikz}
\usepackage{tikz-feynman}
\usetikzlibrary{arrows, positioning, calc, shapes, decorations.markings,
decorations.pathreplacing, math}

\makeatletter
\pgfdeclareshape{myI}
{
    \inheritsavedanchors[from=rectangle]
    \inheritanchorborder[from=rectangle]
    \inheritbackgroundpath[from=rectangle]
    \foreach \i in {center, north, south, east, west, north west, north east,
    south west, south east} {
        \inheritanchor[from=rectangle]{\i}
    }
    \inheritforegroundpath[from=rectangle]

    \backgroundpath
    {
        \def\w{3}
        \def\h{2.5}
        \pgfpathmoveto{\pgfpointorigin}
        \pgfpathlineto{\pgfpoint{0}{\h}}
        \pgfpathlineto{\pgfpoint{\w}{\h}}
        \pgfpathlineto{\pgfpoint{-\w}{\h}}
        \pgfpathmoveto{\pgfpointorigin}
        \pgfpathlineto{\pgfpoint{0}{-\h}}
        \pgfpathlineto{\pgfpoint{\w}{-\h}}
        \pgfpathlineto{\pgfpoint{-\w}{-\h}}
    }
}
\pgfdeclareshape{myH}
{
    \inheritsavedanchors[from=rectangle]
    \inheritanchorborder[from=rectangle]
    \inheritbackgroundpath[from=rectangle]
    \foreach \i in {center, north, south, east, west, north west, north east,
    south west, south east} {
        \inheritanchor[from=rectangle]{\i}
    }
    \inheritforegroundpath[from=rectangle]

    \backgroundpath
    {
        \def\w{2.5}
        \def\h{3}
        \pgfpathmoveto{\pgfpointorigin}
        \pgfpathlineto{\pgfpoint{\w}{0}}
        \pgfpathlineto{\pgfpoint{\w}{\h}}
        \pgfpathlineto{\pgfpoint{\w}{-\h}}
        \pgfpathmoveto{\pgfpointorigin}
        \pgfpathlineto{\pgfpoint{-\w}{0}}
        \pgfpathlineto{\pgfpoint{-\w}{\h}}
        \pgfpathlineto{\pgfpoint{-\w}{-\h}}
    }
}
\pgfdeclareshape{myX}
{
    \inheritsavedanchors[from=rectangle]
    \inheritanchorborder[from=rectangle]
    \inheritbackgroundpath[from=rectangle]
    \foreach \i in {center, north, south, east, west, north west, north east,
    south west, south east} {
        \inheritanchor[from=rectangle]{\i}
    }
    \inheritforegroundpath[from=rectangle]

    \backgroundpath
    {
        \def\w{3}
        \def\h{2.5}
        \pgfpathmoveto{\pgfpoint{\w}{\h}}
        \pgfpathlineto{\pgfpoint{-\w}{-\h}}
        \pgfpathmoveto{\pgfpoint{\w}{-\h}}
        \pgfpathlineto{\pgfpoint{-\w}{\h}}
    }
}
\makeatother

\tikzset{Vertex/.style={regular polygon, regular polygon sides=4, thick,
         minimum size=0.3cm, draw=vertex_fill, fill=vertex_fill, inner sep = -0.1}}
\tikzset{VertexI/.style={myI, ultra thick, minimum height=0.2cm,
         minimum width = 0.2cm, draw=vertex_fill, inner sep = -0.1}}
\tikzset{VertexX/.style={myX, ultra thick, minimum height=0.2cm,
         minimum width = 0.2cm, draw=vertex_fill, inner sep = -0.1}}
\tikzset{VertexH/.style={myH, ultra thick, minimum height=0.2cm,
         minimum width = 0.2cm, draw=vertex_fill, inner sep = -0.1}}

\tikzset{Propagator/.style=thick}
\tikzset{Fermion/.style={Propagator,postaction={decorate,
        decoration={markings, mark=at position .8
            with {\arrow[black,scale=1]{latex}}}}}}
\tikzset{FermionBend/.style={Propagator,postaction={decorate,
        decoration={markings, mark=at position .5
            with {\arrow[black,scale=0.8]{latex reversed}}}}}}
\tikzset{Interaction/.style={Propagator,decorate, decoration=snake}}

\newcommand{\bvec}[1]{\boldsymbol #1}

\newcommand{\dd}{\mathrm{d}}
\newcommand{\DD}{\mathcal{D}}
\newcommand{\psiarg}{\ensuremath{[\psi , \bar\psi]}}

\newcommand{\makeauthor}[2]{\newcommand{#1}[1]{{%
  \sffamily\color{#2}{%
    \bfseries\begingroup\escapechar=-1\edef\x{\endgroup\string#1}\x:%
  } ##1}}}
\makeauthor{\lk}{BurntOrange}
\makeauthor{\jb}{cyan}
\makeauthor{\jh}{NavyBlue}

\newcommand{\captiontitle}[1]{{\bf{}#1}}
\newcommand{\captionlabel}[1]{{\bf{}(#1)}}
\renewcommand{\Re}{\mathrm{Re}}
\renewcommand{\Im}{\mathrm{Im}}
\newcommand{\bzint}[1]{\int_{\mathrm{BZ}}\!\!\dd\bvec{#1}}
\newcommand{\pzint}[1]{\int_{\mathrm{PZ}}\!\!\dd\bvec{#1}}

\newcommand{\bm}[1]{\boldsymbol{#1}}
\newcommand{\onlinecite}[1]{\cite{#1}}

\makeatletter
\def\convertto#1#2{\strip@pt\dimexpr #2*65536/\number\dimexpr 1#1}
\makeatother
\newcommand{\measurewidths}{{%
  \color{red}
  :::
  TEXT \convertto{in}{\the\textwidth} in
  :::
  COLUMN \convertto{in}{\the\columnwidth} in
  :::
  HEIGHT(ex) \convertto{pt}{1ex} pt
  :::
  WIDTH(em) \convertto{pt}{1em} pt
  :::
}}

\usepackage[normalem]{ulem}
\newcommand{\modify}[2]{{\color{purple}\sout{#1}#2}}
\renewcommand{\modify}[2]{#2}

\journalname{Eur. Phys. J. B}

\def\makeheadbox{{%
\hbox to0pt{\vbox{\baselineskip=10dd\hrule\hbox
to\hsize{\vrule\kern3pt\vbox{\kern3pt
\hbox{\bfseries{}Eur. Phys. J. B (2022) 95:65}
\hbox{https://doi.org/10.1140/epjb/s10051-022-00323-y}
\kern3pt}\hfil\kern3pt\vrule}\hrule}%
\hss}}}

\begin{document}

\title{Reference results for the momentum space functional renormalization group}

\newcommand{\equalContribSym}{$\ddagger$}

\author{
  Jacob Beyer\thanksref[\equalContribSym,]{xx}\thanksref{e1, addr1, addr2, addr4}
  \and
  Jonas B.~Hauck\thanksref[\equalContribSym,]{xx}\thanksref{e2, addr2, addr3}
  \and
  Lennart Klebl\thanksref[\equalContribSym,]{xx}\thanksref{e3, addr2, addr3}
}

\thankstext[\equalContribSym]{xx}{These authors contributed equally}
\thankstext{e1}{\email{\href{mailto:beyer@physik.rwth-aachen.de}{beyer@physik.rwth-aachen.de}}}
\thankstext{e2}{\email{\href{mailto:hauck@physik.rwth-aachen.de}{hauck@physik.rwth-aachen.de}}}
\thankstext{e3}{\email{\href{mailto:klebl@physik.rwth-aachen.de}{klebl@physik.rwth-aachen.de}}}

\institute{%
  Institute for Theoretical Solid State Physics,
  RWTH Aachen University, 52062 Aachen, Germany \label{addr1}%
  \and%
  JARA Fundamentals of Future Information Technology,
  52062 Aachen, Germany \label{addr2}%
  \and%
  School of Physics, University of Melbourne, Parkville, VIC 3010, Australia
  \label{addr4}%
  \and%
  Institute for Theory of Statistical Physics,
  RWTH Aachen University, 52062 Aachen, Germany \label{addr3}%
}

\date{Received: 15.01.2022 / Accepted: date}

\abstractdc{
    The functional renormalization group (FRG), an established computational method
    for quantum many-body phenomena, has been subject to a diversification
    in topical applications, analytic approximations and numerical implementations.
    Despite significant efforts to accomplish a coherent standard through
    benchmarks and the reproduction of previous results, no systematic and
    comprehensive comparison \modify{}{of different momentum space FRG
    implementations} has been provided until now.
    While this has not prevented the publication of relevant scientific results
    we argue that established mutual agreement across realizations will
    strengthen confidence in the method.
    To this end we report explicit implementational details and numerical data
    reproduced thrice \modify{in independent FRG implementations}{independently}
    up to machine accuracy.
    To substantiate the reproducibility of our calculations we scrutinize pillar
    FRG results reported in the literature, and discuss our calculations of
    these reference systems.
    We mean to entice other groups to reproduce and establish this set of
    benchmark FRG results thus propagating the joint effort of the FRG
    community to engage in a shared knowledge repository as a reference standard
    for FRG implementations
}
\maketitle
\section{Introduction}
The functional renormalization group (FRG) is one of the most promising methods
for the analysis of low-temperature instabilities of two dimensional materials~
\cite{dlpena_lichtenstein_honerkamp2017,pena-scherer-2017,o_competing_2021,korea_graphene_2019,klebl-kennes-honerkamp-2020,hauck_honerkamp_kennes-2021,classen-honerkamp-scherer-2019,kiesel-thomale-ea-2012,wang-frg-graphene-2012,dlpena_scherer_honerkamp_2014,Classen-2014-phonon,kennes2018strong,honerkamp-honeycomb-2008,raghu-topological-2008,wang-FeAs-2009,wolf2021triplet}
as it offers an unbiased view, not presupposing the existence of certain phase
transitions~\cite{metzner-salmhofer-honerkamp-2012}.
Because the numerics involved in prediction using the FRG are expensive,
the full breadth of possible dependencies \modify{could not (yet) be captured by
any singular realization}{could only be captured by a few implementations of the
two-dimensional single-band Hubbard model on the square
lattice~\cite{tagliavini-ea-2019-multiloop,hille-ea-2020-quantitative}, while a
generalization to non-$SU(2)$ symmetric or multi-orbital models is still
absent}.
This inherent fragmentation has rendered it difficult to directly compare
distinct implementations. Traditionally only qualitative results have been
employed~\cite{lichtenstein_high-performance_2017}.
\modify{Unfortunately various qualitative features produced by the FRG appear to be
robust against a certain degree of programming errors.}{Unfortunately various
qualitative features produced by the FRG, such as the critical scale and leading
instabilities, have (over the course of converging our three distinct
implementations) proven to be robust against a certain degree of programming
error e.g. the exact ordering of indices in subleading diagrams.}
When trying to predict
many-body instabilities of novel systems, certainty over reference calculations
is crucial to assure validity.
Beyond the qualitative level, the aspiration of the method is to become a
quantitative quantum many-body tool, further emphasizing numerical correctness.
The definition of correctness here refers to a certain formal approximation
level, i.e.~a specific truncation of the hierarchy of FRG equations, as detailed
in~Ref.~\onlinecite{metzner-salmhofer-honerkamp-2012}.
When gauging the physical potential of a certain approximation or comparing
various approximations, validity of numerical implementations is mandatory.
It is therefore desirable to introduce a shared knowledge of the
\enquote{proper} results of the tensor contractions intrinsic to most FRG
calculation.
A reproduction of the \emph{exact} numerics yields certainty over prefactors,
signs as well as contraction indices, not ensured when validating using e.g.~the
phase transitions predicted in the square lattice Hubbard model.
\modify{To initiate this opportunity for consistency we provide in the following
numerical results for the effective interaction which were reproduced by three
different, independently developed FRG codes.}{We try to lay the groundwork for
achieving consistency across different codes by providing momentum-space data of
the frequency independent interaction vertex reproduced by three different,
independently developed FRG codes. We choose this focus as a first step towards
a shared knowledge repository, motivated by the application to strongly
correlated states in two-dimensional materials.}

The paper is structured as follows:
We first give a short briefing on the theory and some technicalities of the
\modify{}{momentum-space} FRG \modify{}{and its truncated unity formfactor
expansion}.
Then we introduce the three distinct implementations and elaborate their areas
of application.
We proceed to provide results in the three tests systems we have chosen, in
order to establish agreement with published references.
The penultimate section presents the exact procedure of our comparison, enabling
the reader to compare their own implementation using the provided parameter and
data sets.
Intricacies and specifics regarding various challenges faced during development
are presented in the Appendix.

\section{Method}
\subsection{Functional Renormalization Group}
The derivation of FRG is given in utmost brevity here, we refer the interested
reader to~Refs.~\onlinecite{metzner-salmhofer-honerkamp-2012,Dupuis_review_2021}
for an in-depth discussion. We commence with the partition function of the
interacting system:
\begin{equation}
    \mathcal Z = \int \DD \psi \DD \bar\psi \, e^{-S\psiarg} \,,
\end{equation}
where $\psi, \bar\psi$ are fermionic Grassmann fields and $S$ is the action of
the system given by single-particle and interacting components:
\begin{equation}
    S\psiarg = -(\bar\psi, Q_0 \psi) + S_I\psiarg \,.
\end{equation}
Here $Q_0 = i k_0 - H_0 + \mu = G_0^{-1}$ denotes the inverse of the
bare one-particle Green's function with $H_0$ as the non-interacting part
of the Hamiltonian, $k_0$ a Matsubara frequency and $\mu$ the
chemical potential.

By adding external sources $\eta,\bar\eta$ and Legendre transforming we obtain
the generating functional, the logarithm of which is the generating functional
for connected Green's functions, $\mathcal W[\eta,\bar\eta]$.
The effective action $\Gamma[\psi,\bar\psi]$ is then obtained by another
Legendre transform of $\mathcal W[\eta,\bar\eta]$:
\begin{equation}
    \begin{split}
        & \mathcal W[\eta , \bar \eta]  =
            - \mathrm{ln} \int \DD \psi \DD \bar\psi \,
            e ^{-S \psiarg +(\bar \eta , \psi) +(\eta , \bar \psi)} \,,\\
        & \Gamma \psiarg = (\bar \psi, \eta) + (\bar \eta, \psi)
            + \mathcal W (\bar \eta, \eta) \,.
        \label{eqn:gen_con_Green}
    \end{split}
\end{equation}
We introduce an artificial scale dependence to the bare propagator by means of a
cutoff function: $G^\Lambda_0(k) = G_0(k) \Theta^{\Lambda}(k)$.
The sole requirement on the regulator $\Theta^\Lambda(k)$ is that it vanishes
for high scales ($\Lambda \rightarrow \infty$) and approaches $1$ as $\Lambda
\rightarrow 0$.
This allows a successive integration of different energy scales.
More details on regulators can be found in \cref{ssec:regulators}.

\begin{figure}
    \centering
    \includestandalone[width=\columnwidth]{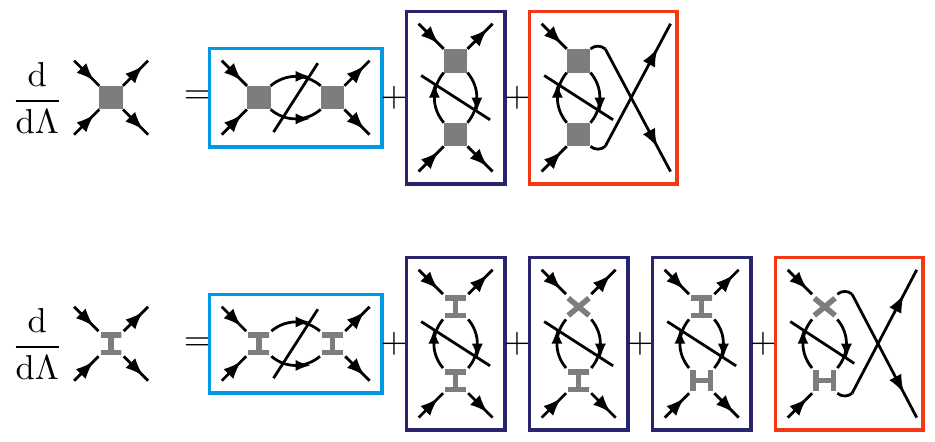}
    \caption{\captiontitle{Diagrammatic representation of the functional
    renormalization group equations.} The lower set of diagrams represent the
    $SU(2)$-symmetric flow equations, where the $SU(2)$-degree of freedom is
    kept constant along the lines of the vertex. The line through the loop
    represents the scale derivative $\dd/\dd\Lambda$. We have already
    indicated the three distinct channels of the FRG flow, differentiated by the
    bosonic transfer momentum: \textcolor{p-channel}{$P$}-,
    \textcolor{c-channel}{$C$}-, and \textcolor{d-channel}{$D$}-channel.}
    \label{fig:diags}
\end{figure}

We can now take a derivative of the effective action with respect to the scale
introduced by means of the regulator and obtain the Wetterich
equation~\cite{Wetterich_1993,Morris_1994}:
\begin{multline}
    \frac{\dd}{\dd\Lambda}\Gamma\psiarg = -(\bar \psi, \dot Q_0^{\Lambda} \psi)
    - \\
    \frac{1}{2}
    \mathrm{tr}\left[\dot{\bm{Q}}_0^\Lambda ({\bm{\Gamma}}^{(2) \Lambda} \psiarg ) ^{-1} \right],
    \label{eqn:flow}
\end{multline}
with 
\begin{equation}
    \bm{Q}^\Lambda_0 =
    \begin{pmatrix}
        Q_0^\Lambda & 0 \\
        0 & -{Q_0^\Lambda}^{\mathrm{T}} \\
    \end{pmatrix}
    \label{eqn:Q_matrix}
\end{equation}
and 
\renewcommand\arraystretch{1.6}
\begin{equation}
    \bm{\Gamma}^{(2) \Lambda} =
    \begin{pmatrix}
        \delta ^2 \Gamma^\Lambda/\delta \bar\psi \delta \psi & \quad
            \delta ^2 \Gamma^\Lambda/\delta \bar\psi \delta \bar \psi \\
        \delta ^2 \Gamma^\Lambda/\delta \psi \delta \psi & \quad
            \delta ^2 \Gamma^\Lambda/\delta \psi \delta \bar \psi \\
    \end{pmatrix} \,.
    \label{eqn:Gam_matrix}
\end{equation}
Expanding in the order of the fields $\psi,\bar\psi$ we obtain an infinite
hierarchy of differential equations with $\Gamma^{\Lambda,(n)}$ dependent on all
even contributions from $\Gamma^{\Lambda,(2)}$ to $\Gamma^{\Lambda,(n+2)}$.
Next let us clarify the truncation level used for the benchmarks presented in
this publication:
We note that higher-order derivatives become decreasingly relevant because they
correspond to multi-electron interactions~\cite{salmhofer-honerkamp-2001-theory}.
Thus, we truncate at $\Gamma^{\Lambda,(4)}$, neglecting all higher order
contributions of the Grassmann field expansion.
Furthermore we neglect the self-energy contribution $\Gamma^{\Lambda,(2)}$ as --
within the scope of this work -- Matsubara frequency dependencies of the vertex
functions are ignored~\cite{honerkamp2003}.
Their inclusion is possible and has been covered extensively \modify{}{for the
2D Hubbard model} in
Refs.~\cite{vilardi2017nonseparable,hille-ea-2020-quantitative,tagliavini-ea-2019-multiloop,reckling-honerkamp-2018,Uebelacker-2021-sffeedback,bonetti_single_2021},
considering the comparison sought here they are however inopportune.
We express these equations as diagrams with $\Gamma^{\Lambda, (4)}=V$%
\footnote{%
  For the sake of simplicity, we do not explicitly distinguish between the
  anti-symmetrized vertex function in the $SU(2)$ symmetric case, usually
  denoted with $V$ and the non-$SU(2)$ vertex function which is usually denoted
  with $\mathcal V$.}
as 4-particle nodes connected by the $G_0$ propagators and their derivative:
$S^{\Lambda}_0(k) = \frac{\dd}{\dd\Lambda} G^{\Lambda}_0(k)$.
The remaining equation is then diagrammatically given in \cref{fig:diags}.
It is insightful to note that up to the truncation of $\Gamma^{(2)}$ and
$\Gamma^{(n\geq6)}$ in the Grassmann field expansion the FRG introduces no
approximations.

To obtain a prediction for the two-particle interaction $V$ independent of the
artificial scale we integrate the ordinary differential flow equation
(\cref{fig:diags}) starting at infinite (large compared to bandwidth) $\Lambda$
and iterating towards $\Lambda=0$.
When encountering a phase-transition, the equations will diverge~
\cite{honerkamp-salmhofer-2001,salmhofer-honerkamp-2001-theory}, making
$\Gamma^{\Lambda,(6)}$ a meaningful contribution and invalidating the above
truncation.
We terminate the integration at this point and examine the final
$V^{\Lambda_\mathrm{c}}$ to determine the ordering associated to the phase
transition.
While this approximation can be improved upon via the inclusion of more
diagrams, e.g.~by applying the more elaborate multiloop-FRG or
Parquet-approximations
\cite{kugler-delft-2018,kugler-delft-2018-2,tagliavini-ea-2019-multiloop,hille-ea-2020-quantitative,eckhardt_truncated_2020},
for ease of comparison and reduction in computational cost it is advantageous to
maintain ease of access.

\subsection{Regulators}
\label{ssec:regulators}
For the calculations in this work two different types of regulators are
used.
The sharp-frequency-cutoff $\Theta^{\Lambda}(k) = \theta(|k_0|-\Lambda)$
and the $\Omega$-cutoff~\cite{husesalm}:
$\Theta^{\Lambda}(k) = |k_0|^2/(|k_0|^2+\Lambda^2)$.
Since we assume a static vertex and neglect self-energy effects we can
analytically calculate all Matsubara frequency summations needed during the
integrations depending on the type of regulator chosen.
Through this we obtain the following loop derivatives for the sharp-cutoff:
\newcommand{\ebl}{\epsilon} %
\newcommand{\eblp}{\epsilon'} %
\begin{align}
\begin{split}
    \dot L_{\pm}^{\Lambda,b_1b_2}(\bvec l,\bvec l')
        = \frac{1}{2\pi}
        &\left[
            \frac{ 1 }{ (i\Lambda - \ebl)(\pm i\Lambda - \eblp)} \right. \\
        +&\left. \frac{ 1 } {(-i\Lambda - \ebl)(\mp i\Lambda - \eblp )}
            \right],
    \label{eqn:Ldot_sharp}
\end{split}
\end{align}
where we used the abbreviated notation $\ebl = \ebl_{b_1}(\bvec l)$ and $\eblp =
\epsilon_{b_2}(\bvec l')$ ($b_1$ and $b_2$ refer to band indices). In the
$\Omega$-cutoff scheme, using the same notation the resulting equations for the
loop derivative are given by:
\begin{equation}
    \dot L_{\pm}^{\Lambda,b_1b_2}(\bvec l, \bvec l') =
    \begin{cases}
      \displaystyle
    \frac{\pm 1/4}{\ebl \mp \eblp}
    \left(
    \frac{\ebl(3\lvert\ebl\rvert + \Lambda)}{(\lvert\ebl\rvert + \Lambda)^3}
    \mp\frac{\eblp(3\lvert\eblp\rvert+\Lambda)}{(\lvert\eblp\rvert+\Lambda)^3}
      \right) \\[1em]
      \displaystyle
    \mp\frac{3\ebl^2-4\lvert\ebl\rvert\Lambda
    -\Lambda^2}{4(\lvert\ebl\rvert+\Lambda)^4}
    \qquad \text{if } \ebl = \eblp \ .
    \end{cases}
    \label{eqn:Ldot_omega}
\end{equation}

In either case the upper ($+$) sign corresponds to the particle-hole loop in
the \textcolor{c-channel}{$C$}- and \textcolor{d-channel}{$D$}-channel diagrams
while the lower ($-$) sign corresponds to the particle-particle loop in the
\textcolor{p-channel}{$P$}-channel diagram.
In the case of the $\Omega$ regulator we neglect the evaluation of the
additional poles of the Matsubara summation which might occur when
$\Lambda=\ebl$, $\Lambda=\eblp$ or $\Lambda=\ebl=\eblp$.
This is justified by calculation of the dispersion only on a discrete grid,
which delegates these cases to be extremely unlikely.

\subsection{Band vs Orbital Calculation}
The above mentioned diagrams (cf.~\cref{fig:diags}) can be evaluated in either
band or orbital space. To illustrate the discrepancies between the two 
we here revert to the
non-diagrammatic representation of the \textcolor{p-channel}{$P$}-channel.
In this channel we can specify $\bvec l'$ to be:
$\bvec l'=\bvec k_1+\bvec k_2-\bvec l = \bvec q - \bvec l$.

We must calculate the loop derivative in band space yielding
$\dot L^{b_1b_2}_{-}(\bvec l,\bvec l')$.
If we desire to integrate the momentum dependence $\bvec l$ using objects
represented in orbital space we need to transform this according to
\begin{multline}
  \dot L_-^{o_1o_2o_3o_4}(\bvec l, \bvec l') =
  \sum_{b_1b_2}
  \dot L_-^{b_1b_2}(\bvec l,\bvec l')
  {u^{o_1b_1}(\bvec l)} {u^{o_2b_2}(\bvec l')} \\
  u^{*o_3b_1}(\bvec l) u^{*o_4b_2}(\bvec l')
  \label{eqn:L_band_to_orb}
\end{multline}
in each flow iteration.
Alternatively we can transform the initial vertex into band space and revert to
orbital space using
\begin{multline}
    V^{b_1b_2b_3b_4}(\bvec k_1,\bvec k_2;\bvec k_3)
    = \sum_{o_1o_2o_3o_4} V^{o_1o_2o_1o_2}(\bvec k_1,\bvec k_2;\bvec k_3) \\
    {u^{*o_1b_1}}(\bvec k_1) {u^{*o_2b_2}}(\bvec k_2)
        u^{o_3b_1}(\bvec k_3) u^{o_4b_2}(\bvec k_1 + \bvec k_2 - \bvec k_3)
        \label{eqn:vertex_to_band}
\end{multline}
and its inverse after the FRG flow. The resulting flow equations in band or
orbital space respectively are given as
\begin{multline}
  \frac{\dd}{\dd\Lambda}\, V_P^{o_1o_2o_3o_4}(\bvec k_1,\bvec k_2;\bvec k_3)
    = \sum_{o'_1o'_2o'_3o'_4} \bzint{l} \\
        \qquad \qquad
        V^{o_1o_2o'_1o'_2}(\bvec k_1,\bvec k_2;\bvec l)
        \dot L^{o'_1o'_2o'_3o'_4}_-(\bvec l,\bvec l') \\
        V^{o'_3o'_4o_3o_4}(\bvec l,\bvec l';\bvec k_3) \,,
        \label{eqn:pdot_orb}
\end{multline}
\begin{multline}
   \frac{\dd}{\dd\Lambda}\, V_P^{b_1b_2b_3b_4}(\bvec k_1,\bvec k_2;\bvec k_3)
    = \sum_{b'_1b'_2} \bzint{l} \\
        \qquad \qquad
        V^{b_1b_2b'_1b'_2}(\bvec k_1,\bvec k_2;\bvec l)
        \dot L^{b'_1b'_2}_-(\bvec l,\bvec l') \\
        V^{b'_1b'_2b_3b_4}(\bvec l,\bvec l';\bvec k_3) \,.
        \label{eqn:pdot_band}
\end{multline}

For convenience we have introduced
\begin{equation}
  \bzint{k} = \pzint{k} = 1 \,,
\end{equation}
the properly normalized integrals over Brillouin zone (BZ) and primitive zone
(PZ).
The computational cost can be reduced by using band space representations of
both vertex and loop, but because $\Gamma^{\Lambda,(4)}$ is not an observable,
this might introduce gauge phases into the calculations.
These can lead to difficulties when interpreting the results and are absent when
calculating in orbital space.
The calculations within this work have therefore been done in orbital
representation, where applicable.

Within the scope of FRG we treat all non-diagonal quantum numbers on equal
footing and use one shared, linearized index $o_i$ to describe spin, orbitals,
sites, etc.
Having dealt with the discrete quantum numbers we now turn our attention to the
more complicated, artificially discretized representation of the continuous
momentum dependencies of the vertex $\bvec k$.

\subsection{Grid-FRG and Refinement}
\label{ssec:grid_refinement}
Since the following section will deal with different methods for the evaluation
of the momentum dependencies we start with an understanding of the fundamental
requirements posed and introduce the basic methods of patching the BZ.

\begin{figure}
    \centering
    \includegraphics[width=\columnwidth]{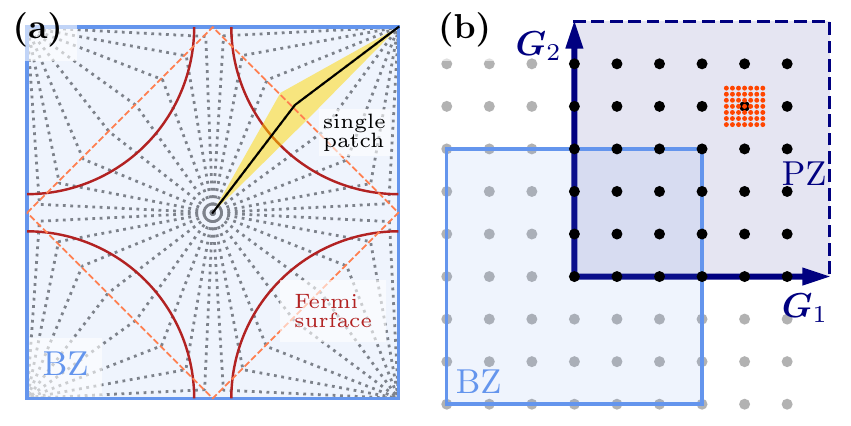}
    \caption{%
      \captiontitle{Meshing schemes for the functional renormalization group
      calculations.} \captionlabel{a} $N$-patch meshing scheme (figure adapted
      from Ref.~\onlinecite{honerkamp_salmhofer_furukawa_rice}), the points are defined at
      the intersections of the Fermi surface (red) with the patch-lines (dashed
      gray). For a single patch, the yellow, filled area is projected to the
      same momentum point.  \captionlabel{b} Bravais momentum mesh in primitive
      zone (PZ) instead of Brillouin zone (BZ) with mesh refinement around one
      selected point shown in orange.%
    }
    \label{fig:npatch_vs_grid}
\end{figure}

The momentum resolution of the functional renormalization group is limited by
two distinct issues:
The lower bound is given by the ability to resolve sharp features of the
fermionic momenta in the loop derivative at low scale.
As these occur in the proximity of the Fermi-surface (FS) this is equivalent to
demanding an accurate representation of said FS within the chosen momentum
scheme.
The upper bound is determined by computational limitations.
Because the memory requirement for saving the vertices -- the largest
objects -- scales proportional to $N_{\bvec k}^3$, raising the fidelity of
momentum discretization is constrained by availability of RAM.

The traditional method of fulfilling these two aspects is to allocate a number
of points along the FS of the system to be used as momentum points for the
vertex calculations [see \cref{fig:npatch_vs_grid}~\captionlabel{a}].
While the vertex would only be defined on the FS, using the assumption that it
is almost constant within the
patch~\cite{honerkamp_doping_2001,honerkamp_salmhofer_furukawa_rice,N-patch-tsai}, the
propagators and their derivatives are evaluated for a set of points spanning the
entire area of the patch.
The values would be averaged to obtain the value of $\dot L$ at the FS patch
point, finely sampling the patches area.
This method is discussed in more detail in
Ref.~\onlinecite{honerkamp_salmhofer_furukawa_rice}.

The immediate issue with this parameterization is the evaluation of the fourth
vertex momentum.
Having constrained $\bvec k_1, \bvec k_2, \bvec k_3$ to points on the FS we have
no capacity to guarantee that $\bvec k_4 = \bvec k_1 + \bvec k_2 - \bvec k_3$
coincides with one of the patch centers.
The assumption is therefore made that this can be projected to the closest
existing point, the center of the patch we land within.
This method, while historically successful in the prediction of fundamental
models~\cite{kiesel-thomale-ea-2012,dlpena_scherer_honerkamp_2014} breaks
momentum conservation of the interaction.
This is especially problematic when attempting to analyze spin-orbit-coupled
(SOC) or multi-site systems where conserving momentum becomes
important due to momentum-spin-locking or momentum-orbit-locking.

The easiest method to restore momentum conservation is the introduction of an
equispaced (\enquote{Bravais}) mesh within the first PZ.
For this case, shown in \cref{fig:npatch_vs_grid}~\captionlabel{b}, the
addition/subtraction of any number of momentum points is assured to be a valid
momentum point.
The obvious issue though is that the majority of points is now far from the FS,
making the resolution of the aforementioned features doubtful.

If we were able to use arbitrarily high momentum resolutions this method would
yield the exact momentum integrations.
It therefore is the base-case for the simulations performed in this paper and
offers the easiest implementation of all two dimensional momentum space FRG
alternatives.
The main challenges to overcome are the memory constrains imposed by the scaling
with momentum resolution and orbital degrees of freedom.
To circumvent this issue we introduce an averaging similar to the one performed
in the evaluation of $N$-patch as
\begin{equation}
  \dot L^{ijmn}_{\pm}(\bvec l, \bvec l') = \sum_{\bvec l_f}\dot L^{ijmn}_\pm(\bvec l
  + \bvec l_f, \bvec l' \pm \bvec l_f),
\end{equation}
where $\bvec l$ is taken from the normal (coarse) fermionic mesh and $\bvec l_f$
is a small (refined) offset relative to this point. We choose the sign to
conserve total momentum in each channel.
The resulting refined grid is shown in
\cref{fig:npatch_vs_grid}~\captionlabel{b}.
Crucially, we employ the refinement only for an averaging within the
calculations of the loop derivative and require no information of the vertex on
the full set of refined momentum points.
As in $N$-patch, we assume the interactions to be sufficiently slowly varying
such that $V(\bvec k_1 + \bvec l_f,\bvec k_2, \bvec k_3) \approx V(\bvec k_1,
\bvec k_2, \bvec k_3) $ holds. This approximation is even more reasonable here
as the maximum distance from the center is greatly decreased.
Some intricacies on the exact implementation of choosing $i,j,m,n$ from orbital
or band space are discussed in \ref{sec:A_refinement} and are omitted here.

\subsection{Truncated Unities}
Building upon the presented basic implementation of grid-FRG we now expand to
further approximations.
The truncated unity extension to the functional renormalization group (TUFRG)
is an approach to reduce the computational complexity introduced using grid-FRG
while maintaining momentum conservation.
Noticing the divergences introduced in the FRG flow will stem from the loop
derivative, we isolate the bosonic momentum $\bvec q$ of $\dot L_\pm(\bvec l,
\pm (\bvec q - \bvec l))$.
For each channel (\textcolor{p-channel}{$P$}, \textcolor{c-channel}{$C$},
\textcolor{d-channel}{$D$}) in \cref{fig:diags} we thus obtain a distinct
transfer momentum:
\begin{subequations}
\begin{align}
  \bvec q_{P} &{}= \bvec k_1 + \bvec k_{2} \,, \\
  \bvec q_{C/D} &{}= \bvec k_1 - \bvec k_{4/3} \,.
\end{align}
\end{subequations}
The remaining weaker fermionic dependencies of vertex and loop are expanded in a
(symmetrized) basis of linearly independent functions $f_m(\bvec k)$ -- called
formfactors.
The choice of these functions is based on the construction of basis functions for
irreducible representations described in
Ref.~\onlinecite{platt-hanke-thomale-2013} with a few improvements and
generalizations detailed in \ref{sec:A_formfactor_generation}.
We project/unproject the flow equations given in \cref{fig:diags} into/from the
formfactor basis using the following operators for each channel $X \in$
\textcolor{p-channel}{$P$}, \textcolor{c-channel}{$C$},
\textcolor{d-channel}{$D$}:
\begin{equation}
\begin{split}
  \hat X [\Phi_X]_{m,n}(\bvec q) &= \bzint{k}\bzint{k'} \\
        &{} \hspace{4em} f_m(\bvec k) f_n^*(\bvec k') \Phi_X(\bvec q,\bvec k,\bvec k') \,, \\
    \Phi_X(\bvec q,\bvec k,\bvec k') &\approx \sum_{m,n} f_m(\bvec k) f_n^*(\bvec k')
        \hat X [\Phi_X]_{m,n}(\bvec q)\,.
\end{split}
\end{equation}
It is instructive to note that with a complete basis of formfactors this is an
exact unitary transformation, but in our case we truncate the basis by
neglecting long-range formfactors.
This is reasonable because the interaction strength declines at long
ranges~\cite{LichtensteinPHD-2018}.
The truncation makes the truncated unity expansion an approximation to the
grid-FRG implementation, but significantly reduces the memory requirements for
the vertices which shrink by a factor of $(N_f/N_{\bvec k})^2$.
This method has been developed in Ref.~\onlinecite{LichtensteinPHD-2018} based
on the earlier work in Ref.~\onlinecite{wang-frg-graphene-2012} and has proven
successful for Hubbard and graphene type
models~\cite{LichtensteinPHD-2018,korea_graphene_2019,o_competing_2021,dlpena_lichtenstein_honerkamp2017}.

When considering the vertex in band space, the gauge phases carried by the
orbital-to-band matrices can not be disregarded.
Thus, the band space vertex cannot be captured accurately in the truncated unity
projections, this approximation thus necessitates orbital space calculations.
TUFRG also has slight symmetry breaking at long coupling ranges for multi-site
models, the details and workaround of this are described in
\ref{sec:A_tu_multisite}.

\subsection{Real Space Truncated Unities}
In models with broken translational symmetry, momentum is not good quantum
number and we thus require a different approach to treat these systems.
The proposal here is to use the real space equivalents of formfactors, which we
will refer to as bonds for the sake of clarity.
In real space the bosonic momentum dependence of each channel is translated to a
dependence on two orbitals (or sites).
The fermionic momenta are then equivalent to the remaining orbital dependencies.
We define the real space bonds $g_{\bvec b_i}(\bvec r_j)$ as Kronecker deltas
spanning the lattice
\begin{equation}
  g_{\bvec b_{i}}(\bvec r_{j}) = \delta({\bvec r_{i} + \bvec b_{i}, \bvec
  r_{j}})\,,
\end{equation}
where $\bvec r_{i}$ is the position of the site $i$ and $\bvec b_{i}$ is a
connection to another site starting at site $i$.
In the case of large unit cells, where momentum is still conserved, it may also
be beneficial to introduce real space truncated unities.
The projections in this mixed space representation then follow
\cref{eqn:P_msproj,eqn:C_msproj,eqn:D_msproj}:
\begin{subequations}
  \begin{multline}
      \label{eqn:P_msproj}
      \hat{P}[V]^{o_1o_3}_{\bvec b_1\bvec b_3}(\bvec q_P)
      =g_{\bvec b_1}(o_2,\bvec k_1) g^*_{\bvec b_3}(o_4,\bvec k_3) \\
      V^{o_1o_2o_3o_4}(\bvec k_1,\bvec q_P-\bvec k_1;\bvec k_3)\,,
  \end{multline}
  \begin{multline}
      \label{eqn:C_msproj}
      \hat{C}[V]^{o_1o_3}_{\bvec b_1\bvec b_3}(\bvec q_C)
      =g_{\bvec b_1}(o_4,\bvec k_1)g^*_{\bvec b_3}(o_2,\bvec k_3) \\
      V^{o_1o_2o_3o_4}(\bvec k_1,\bvec k_1-\bvec q_C;\bvec k_3)\,,
  \end{multline}
  \begin{multline}
      \hat{D}[V]^{o_1o_4}_{\bvec b_1\bvec b_4}(\bvec q_D)= g_{\bvec
      b_1}(o_3,\bvec k_1)g^*_{\bvec b_4}(o_2,\bvec k_4) \\
      V^{o_1o_2o_3o_4}(\bvec k_1,\bvec k_4-\bvec q_D;\bvec k_1-\bvec q_D)\,.
      \label{eqn:D_msproj}
  \end{multline}
\end{subequations}
The generation of the sets of real space bonds and momentum space formfactors
can now be performed following one of two different rules:
For the first, simpler one, we assume that the two sets have no connection,
allowing for a two step projection with separate real- and momentum space
unities.
This decouples the implementations but comes with \modify{an}{a} severe drawback:
For models with more than a single site per unit cell, it is impossible to
include the same interaction orders for each site, thus breaking the rotational
and inversion symmetries as detailed in \ref{sec:A_tu_multisite}.
The second approach remedies these concerns by first creating all real space
bonds on the full lattice with $n_{l_1}\cdot n_{k_1}\times n_{l_2}\cdot n_{k_2}$
elements, where $n_{l_i}$ is the number of real space unit cells and $n_{k_i}$
is the number of momentum points in the direction of basis vector $i$.
Afterwards we perform a Fourier transform of all bonds and obtain
formfactor-bonds of the following general form
\begin{align}
g_{\bvec b_i}(\bvec o_j,\bvec k) &=  e^{-i\bvec k\bvec B_{\bvec b_i}}
 \delta(\bvec r_i +\bvec b_i,  \bvec r_j)\,,
\end{align}
where $\bvec B_{\bvec b_i}$ is the beyond unit cell part of the bond
$\bvec b_i$.
The real space TU was first developed for one dimensional systems starting from
an analysis of the perturbative orders
\cite{bauer-functional-2014,weidinger-bauer-vondelft-2017,markhof_detecting_2018}
and was thereafter re-derived for arbitrary dimension in the context of truncated
unities as we present it here~\cite{in-prep-hauck}.

\section{Implementations}
Having explained the theoretical backdrop for the implementations we now want to
discuss the three distinct FRG codes independently developed by the authors,
sharing minor details of the code-base (input routines, output routines).
Each of the programs will be given a short introduction to establish the
functionalities it covers, its merits and the contrast to the other
implementations.

\subsection{Code \#1: grid-FRG}
As the generation of band structures from \emph{ab initio} simulations is common
practice, it is advantageous to be able to use these as starting point for FRG
calculations.
Thus, this code implements the static four-point FRG equations in either
\emph{band} or \emph{orbital} space and on a regular momentum grid and allows
the study of correlation effects of arbitrary periodic systems if the following
can be provided:
\begin{itemize}
    \item A few isolated low-energy bands of the material in momentum space on
      \modify{on}{} a regular, fine momentum mesh.
    \item Addition and subtraction rules for the momentum mesh.
    \item The $4$-point vertex in any basis that can be connected to the band
        basis by unitary transformation (and thus is of the same dimension). As
        $V$ may be an object of large size, it is possible to supply it on a
        coarse momentum mesh.
    \item The orbital to band transform [Bloch functions $u^{ob}(\bvec k)$] for
        all momentum points. In case the $4$-point vertex is given in band
        space, the matrices fulfill $\hat u(\bvec k) \equiv \mathds{1}$.
\end{itemize}
Since the code is designed to operate on very general models defined only by the
points given above, it is capable of running FRG simulations from e.g. Density
Functional Theory (DFT) or tight-binding (TB) datasets.
The latter can be generated from within the code, with appropriate skeletons
given to make usage simple and efficient.
Furthermore the users are enabled to employ their own, custom momentum space
meshings and thus can in principle use the code for conventional $N$-patch FRG
simulations -- even in the multi-orbital scenario.
We facilitate this by inclusion of an appropriate code skeleton.

The main computational challenge faced by this FRG implementation is memory size
and bandwidth.
The message passing interface (\texttt{MPI}) allows the vertex to be distributed
on a large number of cluster nodes (with the restriction that splitting
constrained to the bosonic momentum index $\bvec q_D$).
This code is memory-bound; simulating e.g. a six band model on a $24\times24$
coarse momentum mesh would take $288$ compute nodes requiring
$\sim140\,\mathrm{GiB}$ memory on each node to store the vertex objects.
For fewer band models these requirements are drastically lowered and allow for
quick and robust operation.

\subsection{Code \#2: TUFRG}
The aim of the TUFRG implementation is to allow fast creation and testing of new
material representations.
This is enabled by removing unnecessary constraints imposed in the creation of
previous frameworks, generalizing the implementation as much as possible, while
extending to include SOC and multi-orbital systems.
The cost introduced by this inclusion has necessitated a focus on performance,
while the decision to generalize for all systems has necessitated some
performance hits.

Due to the generalized nature of this implementation it will be worse in large
unit cells compared to Code~\#3 and will be slower than Code~\#1 for grid-FRG.
It is however an extremely accessible version of FRG and should be both easily
usable and adaptable to uncharted problems.
Furthermore it struggles much less under the memory constraining the
calculations in Code~\#1, a similar calculation as described above would need a
single compute node, \texttt{MPI} is required only for calculation speed.

The framework needs as input a tight-binding model with dispersion and
interaction defined in momentum space as well as a momentum basis.
From this the entirety of the formfactor expansion and TUFRG calculation will be
performed.
As the truncated unities are based heavily on grid-FRG the framework also
provides a basic implementation of grid-FRG.
While this framework is capable of doing the here published grid-FRG
simulations, its primary aspiration is the TUFRG.

\subsection{Code \#3: RS-TUFRG}
The objective of this implementation is to provide a flexible and easy to use
real- and mixed space TUFRG algorithm.
It is optimized for large unit cells, i.e.~between $10$ and a few $1000$ sites
per unit cell which allows for the study of many interesting phenomena, such as
edge properties, disorder effects, behavior in quasicrystalline models and
effects of different boundary conditions.
The models also do not have to be defined on a lattice so that structures such
as Barabasi-Albért networks can be analyzed.
This broad application range of course leads to slight performance
deterioration, thus it is advisable to resort to Code \#1 or Code \#2, for small
unit cells or few band problems. By including all formfactor-bonds, this code can 
effectively perform the grid-RG calculations presented here.

The code itself is designed to be easily extendable with user defined models.
For this it requires at least a definition of the underlying lattice or
graph, a tight binding Hamiltonian, and a distance measure.
It also allows for the inclusion of a single Matsubara frequency per channel
during the flow.
This enables the scaling test, where we check for the correct error scaling
behavior of the interaction when compared to exact diagonalization, as shown
in~\ref{sec:scaling_tests}.

\section{Results -- Connecting to previous publications}
\subsection{Reproducibility: The Hubbard Model}
To bolster confidence in the correctness of our results we first reproduce some
established results of functional renormalization group calculations.
The most basic model -- thoroughly analyzed in
Refs.~\onlinecite{honerkamp-salmhofer-2001,honerkamp_doping_2001,honerkamp_salmhofer_furukawa_rice,honerkamp2003,honerkamp_interaction_2004,honerkamp_ferromagnetism_2004,salmhofer_renormalization_2004,husesalm,Uebelacker-2021-sffeedback,husemann_frequency_2012,LichtensteinPHD-2018,tagliavini-ea-2019-multiloop,hille-ea-2020-quantitative,hille_pseudogap_2020,bonetti_single_2021,schaefer-wentzell-multi-2021}
-- is the Hubbard model for cuprates on a two-dimensional square lattice. We
reproduce the results from
Refs.~\onlinecite{honerkamp-salmhofer-2001,LichtensteinPHD-2018}, the
next-nearest-neighbor hopping phase scan of the repulsive Hubbard model at the
van Hove singularity. \modify{}{Note that the critical scale
$\Lambda_\mathrm{c}$, which is derived from the artificial scale parameter
$\Lambda$, is associated with the temperature of the phase transition and thus
corresponds to a physical observable. Throughout this section, we therefore
effectively compare both the critical temperature and the type of ordering to
previous literature results.}

\begin{figure}%
  \hspace{-0.5em}%
    \begin{tikzpicture}%
      \node at (0,0)%
      {\includegraphics[width=\columnwidth]{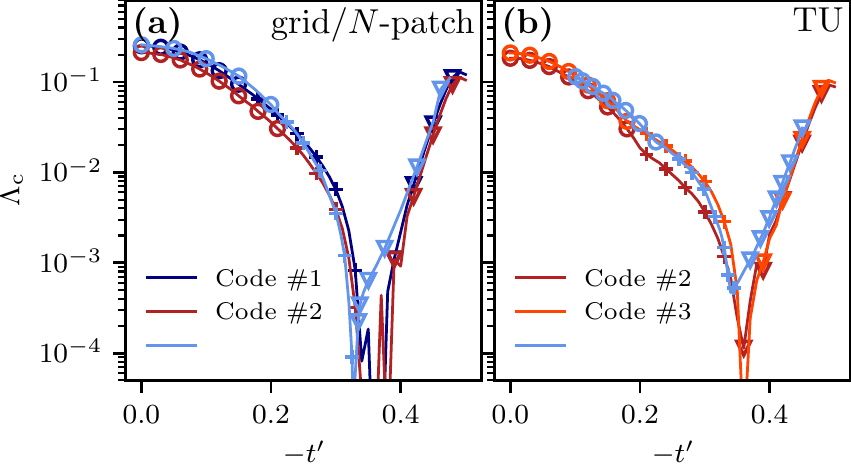}};%
      \node[anchor=south west] at (1.41,-1.32) {\fontsize{7}{1.2}\selectfont%
      Ref.~\cite{LichtensteinPHD-2018}};%
      \node[anchor=south west] at (-2.15,-1.32) {\fontsize{7}{1.2}\selectfont%
      Ref.~\cite{honerkamp-salmhofer-2001}};%
    \end{tikzpicture}%
    \caption{\captiontitle{Square lattice Hubbard model reference results.}
    The plots show the critical scale as a function of next-nearest-neighbor
    hopping $t'$ with the chemical potential fixed at $\mu=-4t'$ to pin the
    Fermi level to the van Hove singularity.
    Additionally, the type of instability is encoded as markers; circles:
    antiferromagnetism, plus signs: $d$-wave superconductivity, triangles:
    ferromagnetism.  The interaction strength is set to $U=3$.
    \captionlabel{a}
    Match of the non-TU results of Codes~\#1 and~\#2 to
    Ref.~\onlinecite{honerkamp-salmhofer-2001}.
    \captionlabel{b}
    TUFRG results from Codes~\#2 and~\#3 matched to the results presented in
    Ref.~\onlinecite{LichtensteinPHD-2018}.
    It should be noted that due to differing details of approximations and
    simulations we do not expect the results to exactly coincide.
    The discrepancies are pronounced in the proximity of the
    transition from superconducting to ferromagnetic phase at $t'\approx-0.35$.
    This region is both very dependent on implementational details as well as
    hard to properly resolve due to the low scales required
    \modify{}{(see~\ref{sec:a_convergence_checks_lambda} for the minimal
    scale allowed by our momentum resolution)}.
    This was already noted in the referred publications but is compounded by the
    different grid discretization chosen in \captionlabel{a} (grid-FRG vs
    N-Patch) leading to the behavior shown.
    }
    \label{fig:Honerkamp-Salmhofer-Compare}
\end{figure}

The non-interacting part of the $t-t'$ Hubbard Hamiltonian is given by:
\begin{equation}
  H_0 = - \sum_{ij,\sigma} t_{ij} \, c_{i,\sigma}^\dagger
  c_{j,\sigma}^{\phantom\dagger}\,,
    \label{eqn:hubbard_h0}
\end{equation}
where the hopping amplitudes are $t_{\braket{i,j}} = 1$ if $i,j$ are nearest
neighbors and $t_{\braket{\braket{i,j}}} = t'$ where $i,j$ are next-nearest
neighbors.
We fix the chemical potential to $\mu=-4t'$ pinning the system to van Hove
filling.
The interacting part of the Hamiltonian is governed by
\begin{equation}
  H_\mathrm{I} = U\, \sum_{i,\sigma} \, n_{i,\sigma} n_{i,\bar\sigma} \,.
    \label{eqn:hubbard_hI}
\end{equation}
where $n_{i,\sigma} = c^\dagger_{i, \sigma} c^{\phantom{\dagger}}_{i,\sigma}$.
\Cref{fig:Honerkamp-Salmhofer-Compare} demonstrates how all three codes
presented in this work reproduce literature in terms of both the critical scales
and the type of instability\modify{}{ (details on the analysis of the effective
vertex at the critical scale are presented in~\ref{sec:A_analysis_of_V_eff})}.
Note that our \modify{}{numerical} implementations differ
\modify{significantly}{} from each other and the reference in terms of the
specifics and methods chosen within the scope of grid-FRG and TUFRG.
To name examples, the type of integrator, cutoff scheme, stopping scale and
termination criterion were \modify{inconsistent}{not explicitly coordinated}.
We observe that the value of $\Lambda_\mathrm{c}$ and the type of instability
close to a phase transition are \modify{very}{} sensitive to these minor
details, but overall the data shows agreement.
The unattainable equivalence in \cref{fig:Honerkamp-Salmhofer-Compare} is a
motivation for the exact comparison given in \cref{sec:benchmarks}.

\subsection{Non-\ensuremath{SU(2)} Systems: The Rashba Model}
\begin{figure}%
  \hspace{-0.5em}
  \begin{tikzpicture}%
    \node at (0,0)%
    {\includegraphics[width=\columnwidth]{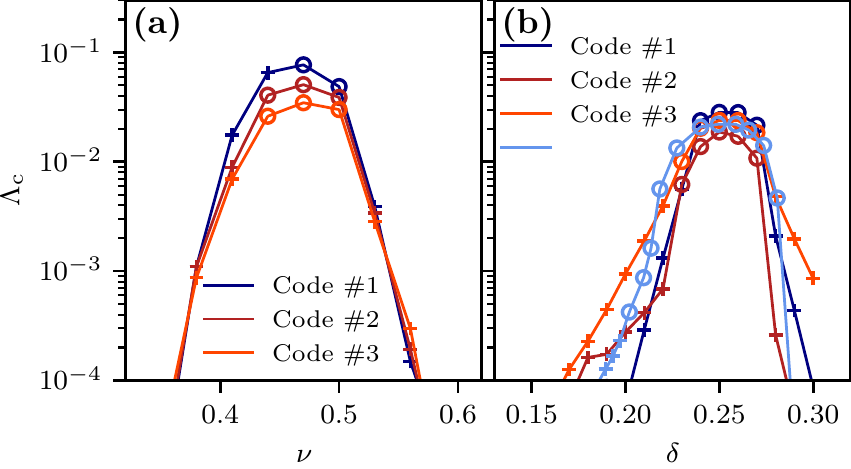}};%
    \node[anchor=south west] at (1.28,0.89-0.33) {{\fontsize{7}{1.2}\selectfont%
    Ref.~\cite{wang-frg-graphene-2012}}};%
  \end{tikzpicture}%
  \caption{\captiontitle{Rashba and graphene model reference results.}
    The plots show the critical scale as a function of filling factor.
    Markers encode the type of ordering;
    circles: antiferromagnetism, plus signs: $d$-wave superconductivity.
  \captionlabel{a} Rashba model. The complete analysis can be found in
    Ref.~\cite{in-prep-rashba}, here we only show internal consistency between
    the three implementations.
    We use the following parameters: $t'=0.15$, $\alpha=0.1$, $U=3$ and the
    filling factor $\nu \in [0.35,0.59]$ with $\nu=0$ ($\nu=1$) corresponding to
    a completely empty (filled) system.
  \captionlabel{b} Graphene model.
    The markers and colors follow the same scheme as in \captionlabel{a}.
    We choose the same parameter set as in
    Ref.~\onlinecite{wang-frg-graphene-2012}: $t'=0.1$, $U=3.6$.
    For ease of comparison we use the doping $\delta=2\nu-1$ instead of the
    filling factor.
    The transition from magnetic to superconducting phase occurs at
    slightly different temperatures than in the reference which however
    employs singular mode FRG.  Graphene is also notoriously resolution
    dependent (momenta and formfactors) in FRG calculations due to
    incommensurate phases~\cite{o_competing_2021} and the multi-site nature
    of the system.  We nonetheless obtain the central results of the
    publication at $\delta\approx1/4$ as well as the transition on either side.
    }
  \label{fig:rashba_graphene_results_compare}
\end{figure}

To confirm all three codes' capabilities extend beyond the $SU(2)$ symmetric
one-band Hubbard model, we introduce a Rashba-$z$ spin-orbit coupling (SOC) term
to the kinetic part of the Hamiltonian breaking the $SU(2)$ symmetry.
It then reads
\begin{equation}
  H_0 =- \sum_{ij,\sigma} t_{ij} \, c_{i,\sigma}^\dagger
  c_{j,\sigma}^{\phantom\dagger}\, - i\alpha\sum_{\braket{ij},\sigma\sigma'}
  \big( \hat{\bvec \sigma} \times \bvec{b}_{ij}
  \big)_z^{\sigma\sigma'}c^{\dagger}_{i,\sigma}
  c^{\phantom\dagger}_{j,\sigma'}\,,
    \label{eqn:rashba_h0}
\end{equation}
where $\alpha$ is the SOC strength, $\hat{\bvec \sigma}$ is the vector of Pauli
matrices, $\braket{ij}$ denote nearest neighbors and $\bvec b_{ij}$ are the
corresponding directed bonds.
Extensive coverage of Rashba-$z$ SOC in the square lattice Hubbard model will be
provided by a publication in preparation \cite{in-prep-rashba} where we employ
FRG and the weak coupling renormalization group~\cite{wolf-2020-rashba,in-prep-schwemmer}.
Here we indicate only the results of a filling phase scan at fixed $\alpha=0.1,
t'=-0.15$ and $U=3$ in
\cref{fig:rashba_graphene_results_compare}~\captionlabel{a}
to demonstrate internal consistency.

\subsection{Multi-band Systems: Graphene}
The third class of models we can check consistent results for are multi-site
unit cell models.
Ref.~\onlinecite{wang-frg-graphene-2012} provides singular mode FRG results for
a tight-binding Hamiltonian of graphene to which we compare the three codes.
The non-interacting part defined on the honeycomb lattice reads
\begin{align}
  H_0 = \sum_{ij,oo',\sigma} t_{ij,oo'} \, c^\dagger_{i,o,\sigma}
  c^{\phantom\dagger}_{j,o',\sigma}\,,
\end{align}
where $o,o'$ are the sites within a unit cell, $i,j$ are unit cell indices and
$t_{ij,oo'}$ the hopping parameters.
We set $t_{ij,oo'}=1$ for nearest-neighbors and $t_{ij,oo'}=t'$ for
next-nearest-neighbors.
For the reproduction of the results in
\cref{fig:rashba_graphene_results_compare}~\captionlabel{b} we choose $t'=0.1$.
The site-dependent Hubbard interaction with $U=3.6$ reads
\begin{align}
  H_\mathrm{I} = U\sum_{i,o,\sigma}\,n_{i,o,\sigma} n_{i,o,\bar\sigma}\,.
\end{align}
Additionally to the results in
\cref{fig:rashba_graphene_results_compare}~\captionlabel{b}, we have reproduced
the critical interaction strength of the half-filled system ($\nu=0.5$) without
next-nearest-neighbor hopping ($t'=0$) to be $U_\mathrm{crit}\approx 2.7t$.
This is in agreement with both Ref.~\cite{SnchezdelaPea:754628} as well as the
expectation that FRG lies between mean-field $U_\mathrm{crit}\approx
2.2t$~\cite{Sorella_1992} and quantum Monte-Carlo $U_\mathrm{crit}=
3.6t-4.0t$~\cite{doi:10.1143/JPSJ.70.1483,paiva-qmc-2005,meng-2010-quantum,sorella_absence_2012}
predictions.

\section{Results -- Benchmark Systems}
\label{sec:benchmarks}
The test systems were chosen to cover a breadth of different aspects of
functional renormalization group calculations in an attempt to reveal common
errors.
The two dimensional square lattice Hubbard model is chosen as the starting point.
We choose the $SU(2)$ symmetric representation which warrants the usage of
the $SU(2)$ symmetric set of flow equations from \cref{fig:diags}.
To also cover the non-$SU(2)$ flow equations it is imperative to include a
non-$SU(2)$ symmetric model.
We therefore also consider the square-lattice Rashba model with spin-orbit
coupling.
While this pair of models should complete the verification of the momentum
dependencies in the contractions calculated during the flow, we want to extend
the benchmarks to include a multi-band model with non-orthogonal lattice
vectors.
This is crucial in finding phase-errors as well as non-periodicities of the
Hamiltonian.
We therefore evaluate and publish data for a honeycomb lattice Hubbard model,
which is a simple model for graphene.
While this is obviously not a complete list of possible models one could show
equivalence for, inconsistencies in this subset should uncover most
inconsistencies that can arise during the implementation of the FRG.
If more models are desirable or advisable the equivalence class will be extended
and the repository~\cite{data} updated appropriately.

\subsection{Square lattice \ensuremath{SU(2)}: The Hubbard Model}
The simplest test case is the square-lattice Hubbard model. Having indicated
that our codes reproduce previously published results we now define parameters
sets for the benchmarks.
The chosen way of integration is grid-FRG -- this allows the greatest
confirmation across the three codes -- other reproduction of data are available
for comparisons upon request.
Be mindful that RS-TUFRG or TUFRG benchmarks can only be supported by a subset
of implementations.

\paragraph{Model:}
To make the test as accessible as possible we shall fix $t=1$, this allows
implementations which use this -- rather common -- scale to implement a
comparative calculation.
The remaining parameters of \cref{eqn:hubbard_h0,eqn:hubbard_hI} are
(arbitrarily) defined to the following nonzero values: $U=3.0$ and $t'=-0.1$.
In the case of the single band $SU(2)$ symmetric Hubbard model the interaction
is constant:
\begin{equation}
  V^{}(\bvec k_1,\bvec k_2,\bvec k_3) = U\,.
\end{equation}
The chemical potential is defined as $\mu=0.5$.
We chose to define the chemical potential rather than the filling to remove the
indirection of its calculation.
This process is inaccurate, especially at the small system sizes discussed here.
\modify{}{As the first step of any momentum space FRG calculation likely is
checking for correctness in the non-interacting part of the Hamiltonian, we
provide band structures along the high-symmetry paths for each of the models
discussed in~\ref{sec:A_band_structures}.}

\begin{figure}
    \centering
    \includegraphics[width=\columnwidth]{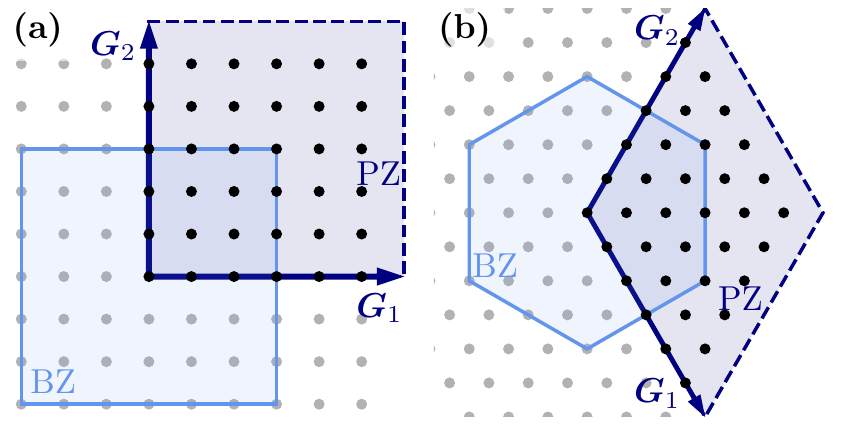}
    \caption{\captiontitle{Momentum space meshing in test cases.}
    \captionlabel{a} Square lattice models (Hubbard and Rashba) have a square
    Brillouin zone (BZ) and a square primitive zone (PZ) \captionlabel{b}
    Triagonal lattice systems (e.g.~graphene) have a hexagonal BZ and a PZ in rhombus
    shape. Both meshes are an equispaced grid along the basis vectors of the
    reciprocal lattices $\bvec G_1$ and $\bvec G_2$. In the hexagonal case we
    can see that the PZ mesh will fill the BZ only after backfolding of points,
    this is intended behavior.}
    \label{fig:meshing}
\end{figure}

\paragraph{Meshing:}
It is essential for exact numerical accuracy that the momentum space is meshed
identically.
We propose the following patching scheme: An equispaced grid of $6$ points along
each edge is laid within the first PZ to include all high-symmetry-points.
36 is a sensible choice for the total number of points because while being small
enough to be quickly calculated it still contains points that are \emph{not}
high-symmetry points.
The resulting mesh can be seen in \cref{fig:meshing}~\captionlabel{a}.
This mesh is used wherever momenta are needed, for the discretization of the vertex as
well as the integration over the loop momentum.
It should be noted that the refinement mentioned in \cref{ssec:grid_refinement}
is not applied in these calculations to reduce sources of potential error.

\paragraph{Flow:}
The most difficult aspect of the calculations to align are the parameters and
calculations involving the integration from high to low scale.
There are many options of integrators, Euler or adaptive, each with a
significantly different update scheme for $\dd/\dd\Lambda$.
Even minute details such as the order of operations (calculating
$\Delta_\Lambda(\Lambda_{n+1})$ or $\Delta_\Lambda(\Lambda_n)$) are relevant.
For this reason we decide to use the simplest possible integrator, constraining
the issues in the numerical analysis:

The integration scheme is a fixed-stepwidth Euler integration which yields
$V(\Lambda_{n+1}) = V(\Lambda_n) + \Delta_\Lambda \frac{\dd V}{\dd
\Lambda}(\Lambda_n)$.
We fix $\Delta_\Lambda = 0.1 \equiv \mathrm{const.}$ and
perform 90 iterations of the FRG-flow starting at $\Lambda = 10$.
We disable all other termination conditions and are thus certain to obtain
$V(\Lambda = 1.0)$.
To avoid misinterpretation, we provide pseudocode for the procedure:
\begin{algorithm}[H]
    \caption{Flow scheme with constant update}
    \begin{algorithmic}[1]
        \State $\Lambda = 10$
        \State $\Delta = 0.1$
        \State Initialize vertex $V = V_0$
        \For {$n=0,1,\ldots,90$}
            \State Evaluate $\dot L (\Lambda)$ with \cref{eqn:Ldot_sharp}
            \State Evaluate $\frac{\dd V}{\dd\Lambda}(\Lambda)$ via
                \cref{fig:diags}
            \State $V\leftarrow V + \Delta\frac{\dd V}{\dd \Lambda}$
            \State $\Lambda\leftarrow\Lambda-\Delta$
        \EndFor
    \end{algorithmic}
\end{algorithm}%
For regulator of the FRG equations we also default to the simplest
possible choice, the sharp cutoff. This is used in all subsequent calculations.

\subsection{Square lattice non-\ensuremath{SU(2)}: The Rashba Model}
The meshing of the square BZ [\cref{fig:meshing}~\captionlabel{a}] as well as
the parametrization of the $\Lambda$ integration are equivalent to the setup
presented for the Hubbard model.
\paragraph{Model:}
We choose the model parameters to ensure broken particle-hole symmetry as well
as broken $SU(2)$ symmetry.
The non-interacting Hamiltonian is defined by \cref{eqn:rashba_h0} while the
interacting part \cref{eqn:hubbard_hI} can be represented as a Hubbard
interaction of electrons of opposite spins:
\begin{multline}
    V^{\sigma_1\sigma_2\sigma_3\sigma_4}(\bvec k_1,\bvec k_2,\bvec k_3) = \\
    U (1-\delta_{\sigma_1\sigma_2})
    \left[
        \delta_{\sigma_1\sigma_3}\delta_{\sigma_2\sigma_4}
        - \delta_{\sigma_1\sigma_4}\delta_{\sigma_2\sigma_3} \right] \,.
\end{multline}
We set the parameters to $t'=-0.1$, $U = 3.0$, $\alpha=0.5$ and $\mu=0.2$.

\subsection{Hexagonal lattice \ensuremath{SU(2)}: The graphene Model}
For the graphene model we choose the parametrization of the FRG-flow to remain
equivalent but need to redefine the BZ meshing
[\cref{fig:meshing}~\captionlabel{b}].

\paragraph{Model:}
The parameters for the calculation of graphene are: $U=3$, $t'=0.1$ and
$\mu=0.2$.
Due to the sublattice structure of graphene the choice of origin in the Fourier
transforms of the Hamiltonian is relevant.
This is detailed in \ref{sec:A_pitfalls}, here it suffices to mention that we
choose the proper gauge which transforms \emph{both} sites of the graphene model
with respect to the origin of the unit cell.
Any other choice of Fourier transform would result in an improper (non-periodic)
gauge of the Hamiltonian.
We specify the interaction in the orbital space of graphene to be a site-local
Hubbard interaction:

\begin{equation}
  V^{o_1o_2o_3o_4}(\bvec k_1,\bvec k_2,\bvec k_3) =
  U\delta_{o_1,o_2}\delta_{o_2,o_3}\delta_{o_3,o_4}
\end{equation}

To be precise we specify the real space lattice vectors ($\bvec a_i$) and
positions of the atoms within the first unit cell ($\bvec x_i$):
\begin{align}
  \bvec a_1 &{}= (\sqrt{3}/2,-1/2,0)^\mathrm{T} &
  \bvec a_2 &{}= (\sqrt{3}/2,1/2,0)^\mathrm{T}\,,\\
  \bvec x_1 &{}= (1/\sqrt{3},0,0)^\mathrm{T} &
  \bvec x_2 &{}= (2/\sqrt{3},0,0)^\mathrm{T}\,.
\end{align}

\paragraph{Meshing:}
The meshing is defined as an equispaced grid of $6 \times 6$ points along the
lines defined by the two reciprocal lattice vectors $\bvec G_1 = (2\pi/\sqrt{3},
-2\pi,0)^\mathrm{T}$ and $\bvec G_2 = (2\pi/\sqrt{3},2\pi,0)^\mathrm{T}$.
This mesh covers the PZ [cf.~\cref{fig:meshing}~\captionlabel{b}] which is
equivalent to any BZ due to the above choice of proper gauge.

Using this meshing ensures an even distribution of the weights to the momentum
points during integrations.
As before this mesh is used for all momenta required in the calculations.

\subsection{TUFRG results}
As Code \#2 as well as Code \#3 are capable of TUFRG calculations we are also
able to generate benchmarks for this approximation.
Equivalence is however much harder to achieve for TUFRG and because the target
audience for the comparison is smaller the results are omitted from this
publication.
Upon request datasets as well as exact descriptions of the procedures involved
will be provided.

\subsection{The Benchmarks}
\subsubsection{Data Availability}
The dataset is publicly available~\cite{data} for comparisons.
Simulation codes will be made available from the corresponding authors upon
reasonable request.
They are expected to be made available in the near future.

\subsubsection{Comparing Instructions}
The comparative data includes the following aspects of the FRG evaluation:
\begin{enumerate}
    \item The maximum contribution to $\Delta_\Lambda\frac{\dd V}{\dd \Lambda}
        (\Lambda)$ by each of the
        \textcolor{p-channel}{$P$}-,\textcolor{c-channel}{$C$}-,
        \textcolor{d-channel}{$D$}-channel diagrams (or the sum of the
        \textcolor{d-channel}{$D$}-channel diagrams for the $SU(2)$ systems)
    \item The maximum vertex element after each iteration of the flow
    \item The final effective vertex $V(\Lambda=1.0)$ at each discretization
      point of the Brillouin zone (and spin/site index for Rashba/Graphene)
\end{enumerate}
The first and second set of data can be compared by calculating the
corresponding values in the trial implementation. The definition of maximum used
is the maximal absolute value of the complex numbers $|z|=\sqrt{\Re(z)^2+\Im(z)^2}$.
Discrepancies in these offer insight into possible faults in singular channels
or prefactors.

The third and most important result to compare is the effective vertex at the
final scale.
Here we offer an array of values (of datatype \texttt{double complex}).
The recommended approach is to sort the reference array as well as the
respective result from the trial code by some metric (i.e. $\max[\Re(z)]$ or
$\max[\Im(z)]$) and compare the sorted arrays.
Note that equivalence can be reached only to computational accuracy
(discrepancies from the resolution of \texttt{double}, rounding, compiler
optimizations, etc. may occur).

While it is also possible to obtain element-wise reproduction, this is dependent
on the order of the discretized momentum points.
Because this stems from the mesh generation scheme as well as the choice of PZ,
neither of which has physical implications, we recommend subverting this
dependency.

\section{Conclusion and Outlook}
This publication set out to rectify the missing link \modify{in the}{for internal}
consistency of \modify{}{momentum space} functional renormalization group
calculations.
We have verified the validity of the implementations by proving agreement with
established results for \modify{}{momentum space} functional renormalization group calculations and
uniformity to unprecedented levels between the three implementations.
This gives us the confidence in the claim that the datasets obtained are
\enquote{correct} in the scope of the specific approximation \modify{}{(i.e.
treating only the four-point vertex in momentum space while neglecting
self-energies and frequency dependencies of the vertex)} of FRG.
We provide these results with the intent of their continued reproduction and
verification by members of the FRG community.
Because the realm of tests can be as vast as the scope of FRG calculations
we invite a continued investment by the community.
The obvious \modify{inclusions}{extensions} include frequency and self-energy
dependent calculations.
Please engage with the authors for required assistance in the reproduction.

Furthermore, the authors are currently working to combine their codes under a
single, versatile \enquote{community code} with a polished, common, easy-to-use
interface.
While this is a major undertaking it is necessary to make the resulting code
accessible to both the FRG community \emph{and} the more general audience of
physicists interested in many-body phenomenæ.

\begin{acknowledgements}
    The GRF (German Research Foundation) is acknowledged for support through RTG
    1995, within the Priority Program SPP 2244 \enquote{2DMP} and under
    Germany’s Excellence Strategy-Cluster of Excellence Matter and Light for
    Quantum Computing (ML4Q) EXC2004/1 - 390534769.
    FRG calculations were performed with computing resources granted by RWTH
    Aachen University under projects rwth0716, rwth0795 and rwth0742.
    We further acknowledge computational resources provided by the Max Planck
    Computing and Data Facility.
    We thank Carsten Honerkamp, Dante M. Kennes and Ronny Thomale
    for their continued mentorship and support. Furthermore we want to
    acknowledge fruitful discussions with Tilman Schwemmer and Stephan
    Rachel.
\end{acknowledgements}

\section{Authors contributions}
Each author was the lead developer on one of the respective implementations.
The authors each acknowledge the other's contributions in the development of
their code in both lines of code and discussions.
All the authors were involved in the preparation of the manuscript and have
contributed equally to this work.

\section*{Appendix}
\appendix

\section{Band structures}
\label{sec:A_band_structures}
\begin{figure*}
  \centering
  \includegraphics[width=\textwidth]{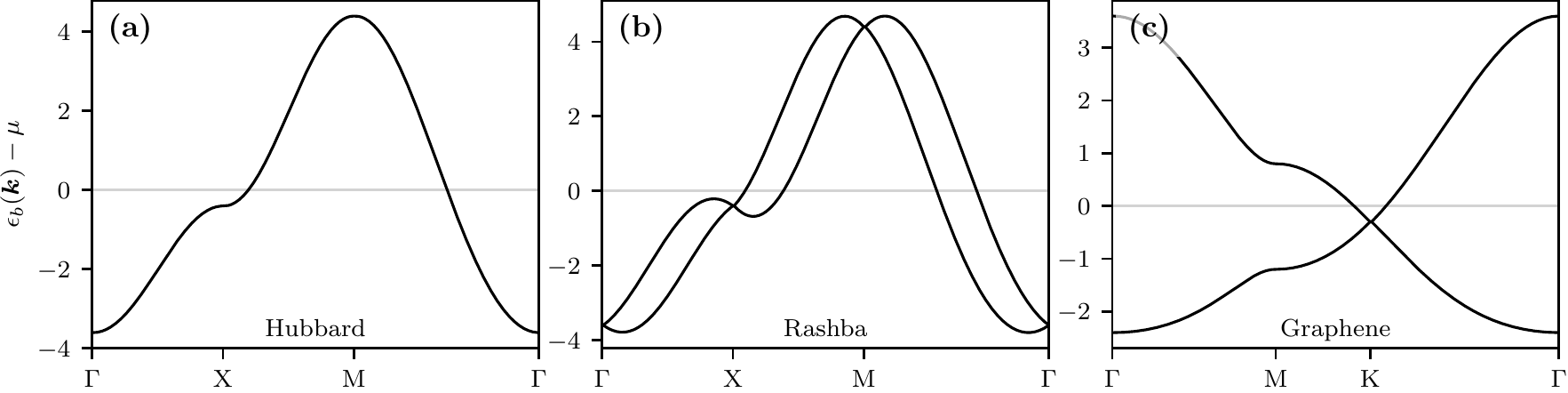}
  \caption{\captiontitle{Band structures} of
    \captionlabel{a} square lattice tight binding model (Hubbard),
    \captionlabel{b} square lattice tight binding model with Rashba-SOC (Rashba)
    and \captionlabel{c} hexagonal lattice tight binding model (graphene), where
    the shorthand notation associates the band structures with the benchmark
    systems as defined in \cref{sec:benchmarks}.
    We note for the Hubbard model the expected saddle point at the X-point.
    The square lattice Rashba band structure has band crossing points at X and M
    while for graphene we draw attention to Dirac cone around the K-point.
    \label{fig:bandstructures}}
\end{figure*}

In Fig.~\ref{fig:bandstructures}, we present the bandstructures of the models
from the three test cases with subtracted chemical potential.
In subfigure~\captionlabel{a}, we show the band structure of the square lattice
Hubbard model with $t = 1$, $t^{\prime} = -0.1$ and $\mu = 0.5$.
In subfigure~\captionlabel{b}, we show the Rashba model as defined in
Eq.~\eqref{eqn:rashba_h0}, with $t = 1$, $t^{\prime} = -0.1$, $\alpha = 0.5$ and
$\mu = 0.2$.
Finally subfigure~\captionlabel{c} shows Graphene with $t = 1$, $t^{\prime} =
0.1$ and $\mu = 0.2$.
These datasets are identical to the benchmark cases, thus reproduction of the band
structures is a good initial test.

\modify{}{%

\begin{figure}
  \modify{}{
  \centering
  \includegraphics[width=\columnwidth]{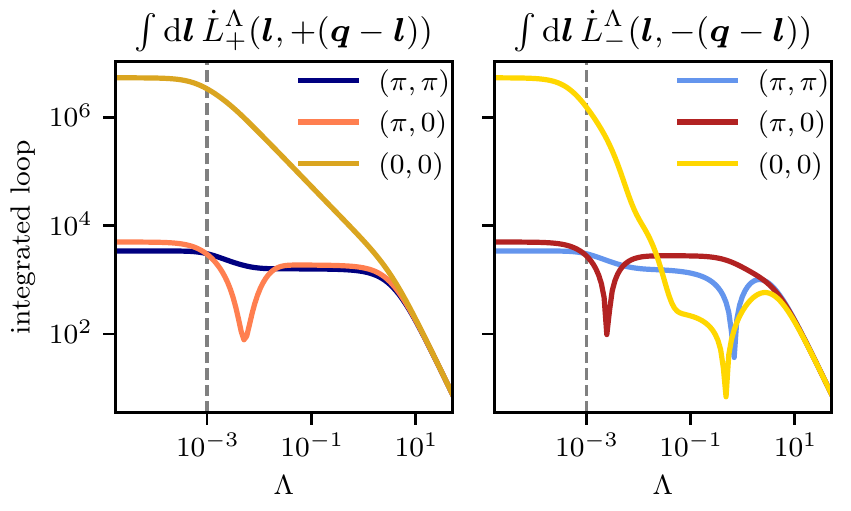}
  \caption{\captiontitle{Integrated particle-hole (left) and particle-particle (right) loops
  $\int\dd\bvec l\,\dot L_\pm^\Lambda(\bvec l,\pm(\bvec q-\bvec l))$ at
  high-symmetry values of $\bvec q$ as a function of scale
  $\Lambda$.} The points $\bvec q$ are indicated in the legends. Note that at
  $\Lambda=10^{-3}$, we add a dashed gray emphasizing the minimum value of
  $\Lambda$ for which the loops can be considered to be converged. Below this
  scale, the spacing of eigenenergies (for the discretization we chose) is
  larger than $\Lambda$ leading to a divergence that is cut off.}
  \label{fig:loop-integrated}
  }
\end{figure}

\section{Minimum Scale and Momentum Resolution}
\label{sec:a_convergence_checks_lambda}
The oscillatory behavior at low scales seen in the critical scale as a function
of $t'$ in the grid-FRG results for the Hubbard model (see
\cref{fig:Honerkamp-Salmhofer-Compare}~\captionlabel{a}) implies that for
$\Lambda<10^{-3}$, the momentum resolution we employed ($20\times20$ coarse
points and $25\times25$ fine points) cannot resolve smaller scales.
To support this, we show the (momentum $\bvec k$) integrated
particle-particle and particle-hole loops at high-symmetry transfer momenta as a
function of $\Lambda$ obtained from the grid-FRG simulation in
\cref{fig:loop-integrated}. We specifically chose the value $t'=-0.35$ to be
consistent with \cref{fig:Honerkamp-Salmhofer-Compare}. As expected, the loops
cannot be resolved to scales smaller than $\Lambda<10^{-3}$, as indicated by the
dashed gray lines.

We want to further highlight the fact that these loop integrals are in fact the
same as the onsite formfactor projection of the TUFRG loops. The non-oscillatory
behavior of the TUFRG codes in
\cref{fig:Honerkamp-Salmhofer-Compare}~\captionlabel{b} as well as the small
kink at around $t'=-0.2$ can be explained by the fact that long-range
interactions cannot fully be captured in TUFRG when they are present in
multiple channels. Therefore, at the phase boundaries, the truncation employed
leads to slightly different behavior than in the grid-FRG case -- the momentum
resolutions employed in the integrations are similar.
}

\section{Analysis of Effective Vertex}
\label{sec:A_analysis_of_V_eff}
Having obtained the effective interaction $V^{\Lambda_\mathrm c}$ from the
evaluation of the flow
equations we list useful
strategies to determine types of phase transitions.
\subsection*{Element Picking}
The simplest method for phase identification is element picking.
Here we consider the elements known to drive different phases and compare their
values, choosing the biggest magnitude as the phase.
This works very well for systems with a finite set of known and simple phases
such as the Hubbard model. We can identify all three phases in
\cref{fig:Honerkamp-Salmhofer-Compare} by means of this.

For the superconducting phase we can consider the element at \modify{$\bvec k_1-\bvec
k_2 = 0$}{$\bvec k_1+\bvec k_2=0$}, using the remaining momentum dependencies to distinguish that we
encounter a $d$-wave superconductor. In the truncated unities approximation this
becomes even more powerful as the other momenta are already expanded into the
formfactor basis.
For the antiferromagnetic phase we consider the momentum transfer
$\bvec k_1-\bvec k_3 = (\pi, \pi)$. As this drives the
\textcolor{c-channel}{$C$}-channel diagram we can
assume a spin-density-wave with wave vector $(\pi,\pi)$ which corresponds -- on
the square lattice -- to antiferromagnetism.
Similarly we can decipher ferromagnetism by examining the vertex at the momentum
transfer $\bvec k_1-\bvec k_3 = (0,0)$, a spin-density-wave with this
wave vector corresponds to a ferromagnetic phase.

\subsection*{Susceptibilities}
For particle-hole instabilities (spin-density waves and charge-density waves in
the $SU(2)$ case), it is instructive to study the (crossed) particle-hole
susceptibility using the vertex given at the end of the FRG flow projected to
the corresponding channels.~\cite{klebl2021moire} To efficiently calculate these, we define the
Fermi particle-hole loop at scale $\Lambda$ as
{ \newcommand{\epschan}{\epsilon^{\phantom{\prime}}}
  \newcommand{\kchan}{\bvec{k}^{\phantom{\prime}}}
  \newcommand{\qchan}{\bvec{q}^{\phantom{\prime}}}
  \begin{multline}
    L_{f,+}^{\Lambda,o_1o_2o_3o_4}(\bvec q_D,\bvec k_D)
    = \sum_{b_1b_2} u^{o_1b_1}(\kchan_D)
    u^{*o_3b_1}(\kchan_D) \\
    \qquad
    u^{o_2b_2}(\bvec k_D')
    u^{*o_4b_2}(\bvec k_D')\, \big[f\big(\epschan_D/\Lambda\big)
      - f\big(\epsilon_D'/\Lambda\big)\big] \\
      \big[ \epschan_D - \epsilon_D' \big]^{-1}\,,
    \label{eqn:fermi-loop-d}
  \end{multline}
  with the Fermi function $f(x)=\big(1+e^x\big)^{-1}$
    and $\epschan_X = \epsilon_{b_1}(\kchan_X),
    \epsilon_X' = \epsilon_{b_2}(\bvec k_X')$.
    From this, we calculate the four-point susceptibility as
  \begin{multline}
      \chi_D^{o_1o_2o_3o_4}(\bvec q_D)
      = \sum_{\substack{
          \bvec k_D^{\phantom{\prime}}\bvec k_D^\prime \\
          o_1'o_2'o_3'o_4'}}\,
      L_{f,+}^{\Lambda, o_1o_2o_1'o_2'}(\qchan_D,\kchan_D)\, \\
      \qquad
    V_D^{\Lambda,o_1'o_2'o_3'o_4'}(\qchan_D, \kchan_D, \bvec k_D')\,
      L_{f,+}^{\Lambda,o_3'o_4'o_3o_4}(\qchan_D,\bvec k_D')\,.
  \end{multline} }

\subsection*{Linearized Gap Equation}
For \textcolor{p-channel}{$P$} instabilities, it is convenient to solve a
linearized gap equation to obtain details about the system's leading ordering
tendencies.~\cite{klebl2021moire} We define the \textcolor{p-channel}{$P$}-channel Fermi loop at scale
$\Lambda$ as
{ \newcommand{\epschan}{\epsilon^{\phantom{\prime}}}
  \newcommand{\kchan}{\bvec{k}^{\phantom{\prime}}}
  \newcommand{\qchan}{\bvec{q}^{\phantom{\prime}}}
  \begin{multline}
    L_{f,-}^{\Lambda, o_1o_2o_3o_4}(\bvec q_P,\bvec k_P)
        = \sum_{b_1b_2} \, u^{o_1b_1}(\kchan_P)
    u^{* o_3b_1}(\kchan_P) \\
    \qquad
    u^{o_2b_2}(\bvec k'_P)
      u^{*o_4b_2}(\bvec k'_P)\, \big[f\big(-\epschan_P / \Lambda\big)
      - f\big(\epsilon_P' / \Lambda\big)\big] \\ \big[
        \epschan_P + \epsilon_P' \big]^{-1}\,.
  \end{multline} }
We can then proceed to define a superconducting linearized gap equation as
\begin{multline}
  \lambda\,\Delta_P^{o_1o_2}(\bvec k) = \sum_{\bvec
  k'o_3o_4o_1'o_2'}
  V_P^{\Lambda,o_1o_2o_3o_4}(\bvec q_P=0,\bvec k,\bvec k')\, \\
  L_{f,\pm}^{\Lambda,o_3o_4o_1'o_2'}(\bvec q_P=0,\bvec k')\,
  \Delta_P^{o_1'o_2'}(\bvec k')\,.
  \label{eqn:frg-eigen}
\end{multline}
The eigenproblem in \cref{eqn:frg-eigen} is in general non-hermitian and thus
numerically unstable. Therefore, we instead solve the following singular value
decomposition:
\begin{equation}
    \hat{V}_P^\Lambda \hat{L}_{f,-}^\Lambda =
    \hat{U}\,\hat{\Sigma}\,\hat{V}^\dagger\,.
\end{equation}
The right singular vectors $\hat V$ are gap functions projected to the Fermi
surface (with \enquote{temperature} broadening set by $\Lambda$) and the left
singular vectors do not include the Fermi surface projection and display the
gap's symmetry. In the case of non-$SU(2)$ and two-site systems, it is
instructive to transform the superconducting (\textcolor{p-channel}{$P$}) gap
to singlet [$\psi(\bvec k)$] and triplet [$\bvec d(\bvec k)$]
space~\cite{sigrist1991phenomenological,smidman2017superconductivity}:
\begin{equation}
    \hat\Delta(\bvec k) = i\,\big[ \hat\sigma_0\psi(\bvec
    k) + \hat{\bvec \sigma}\cdot\bvec d(\bvec k) \big]\,\hat{\sigma}_y,
\end{equation}
with $\hat{\bvec \sigma}$ Pauli matrices and $\hat\sigma_0$ the identity matrix.

\section{Pitfalls during evaluations}
\label{sec:A_pitfalls}
During the production of the results for each model we encountered numerous
possible pitfalls which can lead to different results in the test set. The
first and most important one is to make sure that the models are defined
exactly as above. Other common mistakes that occur are using different sign
conventions or wrong input parameters, so if the results do not match this
should be the first suspect. Otherwise the following section may help
identify problems in the comparison. Most of these pertain to general problems
which should be taken into account when implementing an FRG-code.

\subsection*{Signs and prefactors of the diagrams}
\begin{figure}
  \modify{}{
  \centering
  \includegraphics[width=\columnwidth]{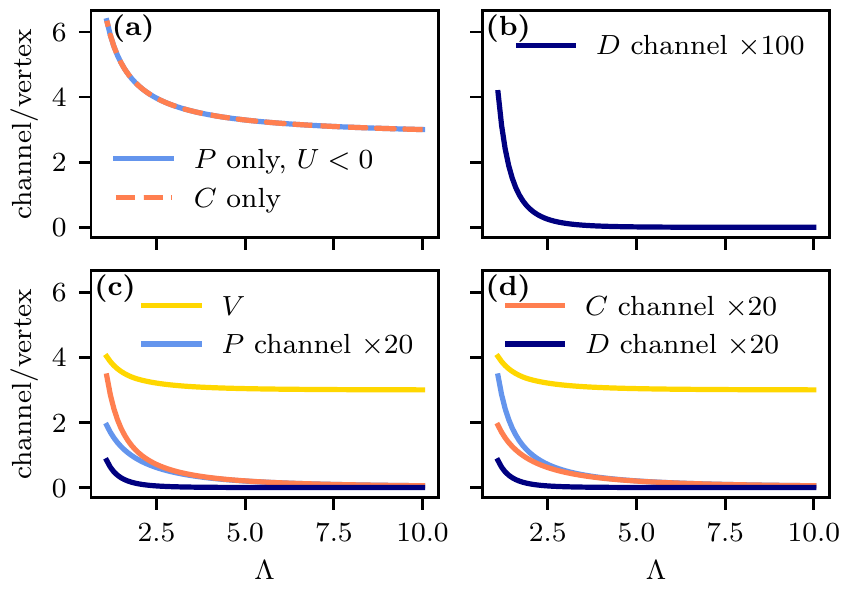}
  \caption{\captiontitle{Graphical representation of test system properties.}
  For all cases, we used the test system parameters, except for $\mu=0$, $t'=0$
  such that the symmetries hold. \captionlabel{a} Vertex maximum as a function
  of scale in single channel flow in $P$ for $U=-3$ (blue) and $C$ for $U=3$.
  The two lines are the same (point~\ref{ssec:su2diag:enum:pc} in the text).
  \captionlabel{b} Full repulsive flow, $D$ channel contribution that is
  identically zero at first step. (cf. point~\ref{ssec:su2diag:enum:dchan0})
  \captionlabel{c} Full repulsive flow, all channels and vertex. (cf.
  point~\ref{ssec:su2diag:enum:attract}) \captionlabel{d} Full attractive flow
  ($U=-3$) with all channels and vertex. (cf.
  point~\ref{ssec:su2diag:enum:repulse})}
  \label{fig:graphical-signs-and-prefactors}
  }
\end{figure}

One of the main issues with functional renormalization group calculations is
fixing the prefactors (signs and values) within the flow equations
(cf.~\cref{fig:diags}). Here we want to list some strategies
which can be employed in order to obtain the correct values for the special case
of the square lattice Hubbard model at half filling (and $t'=0$).
\modify{}{A graphical representation for the $SU(2)$ symmetric rules is shown in
\cref{fig:graphical-signs-and-prefactors}.}

\subsubsection*{Non-\ensuremath{SU(2)} diagrams}
The following rules apply to the three diagrams of the non-$SU(2)$ equation:
\begin{itemize}
    \item For an initially attractive interaction, the absolute value maximum of
        the \textcolor{p-channel}{$P$}-channel contribution must increase over
        the flow (for the first steps).
    \item Using an initially repulsive interaction, the
        \textcolor{c-channel}{$C$}- and \textcolor{d-channel}{$D$}-channel
        contributions to the effective vertex must increase over the flow
        (for the first steps).
    \item The \textcolor{c-channel}{$C$}- and \textcolor{d-channel}{$D$}-channel
      contributions must be equivalent up to a reordering of the orbital and
      momentum indices and sign including the prefactors in the flow equations.
\end{itemize}

\subsubsection*{\ensuremath{SU(2)} diagrams}
\begin{enumerate}
    \item When calculating the single-channel flow in
        \textcolor{c-channel}{$C$}- and \textcolor{p-channel}{$P$}-diagrams the
        results must be equal when inverting $U$.
        \label{ssec:su2diag:enum:pc}
    \item The total value of the \textcolor{d-channel}{$D$}-channel contribution
        must be zero in the first step.
        \label{ssec:su2diag:enum:dchan0}
    \item For an initially repulsive interaction, the
      \textcolor{c-channel}{$C$}-channel contribution to the effective vertex must increase over
      the flow (for the first steps).
        \label{ssec:su2diag:enum:attract}
    \item When calculating an attractive interaction in
        \textcolor{p-channel}{$P$}-channel the maximum of the effective vertex
        must increase over the flow (for the first steps).
        \label{ssec:su2diag:enum:repulse}
\end{enumerate}

\subsection*{BZ-periodicity of systems with basis}
Here we want to address the periodicity of the Hamiltonian in the case of
existing sublattice structure.

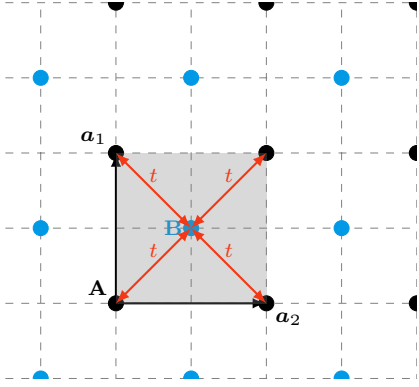
\begin{figure}[ht]
  \centering
    \begin{tikzpicture}[scale=0.5]
    \coordinate (Origin)   at (0,0);

    \clip (-3,-2) rectangle (8,8); 
    \pgftransformcm{1}{0}{0}{1}{\pgfpoint{0cm}{0cm}}
    \coordinate (Bone) at (0,4);
    \coordinate (Btwo) at (4,0);
    \draw[style=help lines,dashed] (-8,-8) grid[step=2] (8,8);

    \foreach \x in {-4,-2,...,4}{
      \foreach \y in {-4,-2,...,4}{
        \node[draw,circle,inner sep=2pt,fill] at (2*\x,2*\y) {};
        \node[draw=p-channel,circle,inner sep=2pt,fill=p-channel] at (2*\x+2,2*\y+2) {};
      }
    }
    \node[above left] at (0,0) {\textbf{A}};
    \node[left, text=p-channel] at (2,2) {\textbf{B}};
    \draw [thick,-latex] (Origin)
        -- (Bone) node [above left] {$\bvec a_1$};
    \draw [thick,-latex] (Origin)
        -- (Btwo) node [below right] {$\bvec a_2$};
    \filldraw[fill=gray, fill opacity=0.3, draw=none] (Origin)
        rectangle ($(Bone)+(Btwo)$);
      \draw [thick,latex-latex, c-channel] (Origin) -- ($0.5*(Bone)+0.5*(Btwo)$)
      node[midway,above] {$t$} ;
      \draw [thick,latex-latex, c-channel] (Bone) -- ($0.5*(Bone)+0.5*(Btwo)$)
      node[midway,above] {$t$} ;
      \draw [thick,latex-latex, c-channel] (Btwo) -- ($0.5*(Bone)+0.5*(Btwo)$)
      node[midway,above] {$t$} ;
      \draw [thick,latex-latex, c-channel] ($(Bone)+(Btwo)$) -- ($0.5*(Bone)+0.5*(Btwo)$)
      node[midway,above] {$t$} ;
  \end{tikzpicture}
  \caption{\captiontitle{Example lattice with sublattice structure with
  nearest-neighbor hopping $t$ from $B$ to $A$ site.}}
  \label{fig:sublattice_structure}
\end{figure}

Fourier transforming into momentum space there are two options for handling the
positions of the orbitals. We can either use the actual positions in the
transformation or we map all sites onto the origin and Fourier transform with
respect to this. Using the first option introduces non-periodicities. We will
illustrate this using a simple two site model:
\begin{equation}
\begin{split}
    H_0 = t \sum_{\braket{ij}}
    &\left(
    c^{\dagger}_{i,A}c^{\phantom{\dagger}}_{j,A}
    + c^{\dagger}_{i,B}c^{\phantom{\dagger}}_{j,B}
    + c^{\dagger}_{i,A}c^{\phantom{\dagger}}_{i,B} \right. \\
    &\quad \left.+ c^{\dagger}_{i,B}c^{\phantom{\dagger}}_{i,A}
    + c^{\dagger}_{i,A}c^{\phantom{\dagger}}_{j,B}
    + c^{\dagger}_{j,B}c^{\phantom{\dagger}}_{i,A}
    \right) \,.
\end{split}
\end{equation}
We define the Fourier transforms with respect to the origin as follows:
\begin{equation}
    c^{\phantom{\dagger}}_{i,A}
    =
    \bzint k \, e^{i\bvec k\bvec{R}_i} c^{\phantom{\dagger}}_A
\end{equation}
and
\begin{equation}
    c^{\phantom{\dagger}}_{i,B}
    =
    \bzint k \, e^{i\bvec k\bvec{R}_i} c^{\phantom{\dagger}}_B
\end{equation}
resulting in the non-interacting Hamiltonian of the form
\begin{multline}
  H_0 = t \bzint k\, \bigg\{
        c^{\dagger}_{A}(\bvec k)c^{\phantom{\dagger}}_{A}(\bvec k)
      + c^{\dagger}_{B}(\bvec k)c^{\phantom{\dagger}}_{B}(\bvec k)
      + c^{\dagger}_{A}(\bvec k)c^{\phantom{\dagger}}_{B}(\bvec k) \\
      \qquad
      {} + c^{\dagger}_{B}(\bvec k)c^{\phantom{\dagger}}_{A}(\bvec k)
      + \hspace{-1em} \sum_{\substack{\bvec{v}\in \{\bvec a_1,\bvec a_2, \\
    \bvec a_1+\bvec a_2\}}}\hspace{-1em}
        e^{i\bvec k\bvec{v}}
        \Big[
        c^{\dagger}_{A}(\bvec k)c^{\phantom{\dagger}}_{B}(\bvec k) \\
      {} + c^{\dagger}_{B}(\bvec k)c^{\phantom{\dagger}}_{A}(\bvec k)
        \Big] \bigg\}\,,
\end{multline}
which is a perfectly $2\pi$ periodic form of the non-interacting Hamiltonian.
We shall call this gauge a \enquote{proper} gauge. Now to study the effects of
an improper gauge, where we use the actual positioning of the sublattice within
the unit cell to define the improper gauged Fourier transforms:
\begin{equation}
    c^{\phantom{\dagger}}_{i,A}
    =
    \bzint k\, e^{i\bvec k\bvec{R}_i} c^{\phantom{\dagger}}_A
\end{equation}
and
\begin{equation}
    c^{\phantom{\dagger}}_{i,B}
    =
    \bzint k\,
    e^{i\bvec k(\bvec{R}_i+\frac{\bvec a_1}{2}+\frac{\bvec a_2}{2})}
    c^{\phantom{\dagger}}_B \,.
\end{equation}
Using these expressions to Fourier transform $H_0$ we can see that we collect some
phase-terms which are not $2\pi$ periodic in $\bvec k$:
\begin{multline}
  H_0 = t \bzint k\, \bigg\{
        c^{\dagger}_{A}(\bvec k)c^{\phantom{\dagger}}_{A}(\bvec k)
      + c^{\dagger}_{B}(\bvec k)c^{\phantom{\dagger}}_{B}(\bvec k)
      + c^{\dagger}_{A}(\bvec k)c^{\phantom{\dagger}}_{B}(\bvec k) \\
      \qquad
      + c^{\dagger}_{B}(\bvec k)c^{\phantom{\dagger}}_{A}(\bvec k)
      + \hspace{-1em}\sum_{\substack{\bvec{v}\in \{\bvec a_1,\bvec a_2, \\
    \bvec a_1+\bvec a_2\}}}\hspace{-1em} e^{i\bvec k\bvec{v}}
        e^{i\bvec k(\frac{\bvec a_1}{2}+\frac{\bvec a_2}{2})}
        \Big[
        c^{\dagger}_{A}(\bvec k)c^{\phantom{\dagger}}_{B}(\bvec k) \\
      + c^{\dagger}_{B}(\bvec k)c^{\phantom{\dagger}}_{A}(\bvec k)
        \Big] \bigg\} \,.
\end{multline}
We can therefore no longer evaluate $H_0$ after backfolding into the first BZ,
instead we need to evaluate them in higher BZs.

While in theory this is a seemingly natural gauge, the periodicity of the
Hamiltonian is required for the momentum meshing used in this work. Furthermore
using an improper gauge will change the topological properties of the model
system. The definition of a $2\pi$ periodic gauged representation is therefore
of paramount importance.

\subsection*{Handedness of Coordinate systems}
While it may be obvious we want to briefly discuss the handedness of the momentum
lattice. When defining a basis of three vectors spanning space we
should define the third vector such that the span product $v = (\bvec
x\times\bvec y)\cdot\bvec z > 0$. This is called a right-handed coordinate system.

When defining the real basis $\bvec a_i$ of the system instead we take care that
it is right handed, the momentum lattice vectors $\bvec G_i$ can then be
obtained via the transformations:
\begin{equation}
    [\bvec G_1, \bvec G_2, \bvec G_3] = 2\pi\left([\bvec a_1, \bvec a_2, \bvec
    a_3]^{-1}\right)^\mathrm{T} \,.
\end{equation}

\section{Refinement and Symmetries}
\label{sec:A_refinement}
Using the refinement to subsum the dependencies within the area associated
with a mesh-point we run into issues which slightly break the symmetry of the
Hamiltonian. For clarity we will refer to the center the refinement collapses
into as coarse point.

\subsubsection*{Definition of symmetric refinement}
\begin{figure}
    \centering
    \includegraphics[width=\columnwidth]{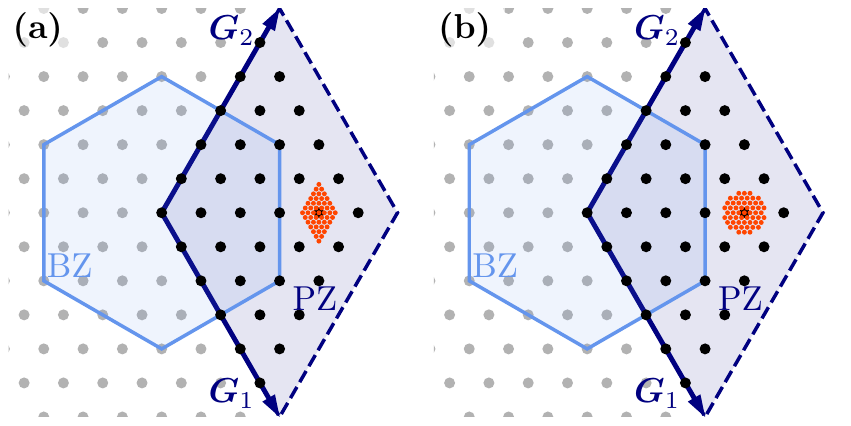}
    \caption{\captiontitle{Refinement under symmetries.}
    \captionlabel{a} Intuitive but incorrect definition of the refined mesh
    atop the coarse mesh. As can clearly be seen under a rotation of $C_3$ the
    indicated coarse points transform into each other, a rotation of the
    refinement however renders it incommensurate.
    \captionlabel{b}
    Proper definition of the refinement area as the Wigner-Seitz cell around
    the coarse point. It is apparent that these points will respect all
    symmetry transformations of the coarse mesh.  }
    \label{fig:A_nonsymmetric_refinement}.
\end{figure}

When defining the refined mesh points atop the coarse mesh points some attention
is needed to obtain a set invariant under the symmetry operations of the coarse
mesh.
For an illustration of the set let us consider a hexagonal BZ into which we
have introduced the coarse mesh via the methods discussed in
\cref{ssec:grid_refinement}. 
Defining the refined mesh as a scaled-down version of the coarse mesh would
result in a mesh which is not invariant under the symmetry transformations, this
is blatantly obvious from \cref{fig:A_nonsymmetric_refinement}.
The proper method employed in the implementations discussed here is to define a
refined mesh of multiple times the size of the refined area size and reduce the
points to the set which is closest to the coarse center point [Wigner-Seitz like
construction, cf. \cref{fig:A_nonsymmetric_refinement}~\captionlabel{b}].
Depending on the system's lattice, it may be needed to double-count some of the
refined points and introduce non-unitary integration weights.

\subsection*{Orbital-Band-Transformation symmetry breaking}
Even when using a properly symmetric refined mesh there are remaining symmetry
issues in the orbital-to-band-transformations:
We assume a model system with momentum-dependent orbital-orbital symmetry
relations (such as the mapping of graphene-sites under rotation).
When we introduce refinement we have two options of calculating the averaged
non-refined loop:

\begin{itemize}
    \item Calculate the average in band space, then transform into orbital space
    using the transformation matrices of the coarse points.
    This is subject to the band-crossing problem presented in the
    paragraph \enquote{Band Crossing Problem}
    \item Transform each refined point into orbital space and average in orbital
    space.
    This is subject to the symmetry breaking of refinement presented the paragraph
    \enquote{Summed Symmetries}.
\end{itemize}

\subsubsection*{Band Crossing Problem:}
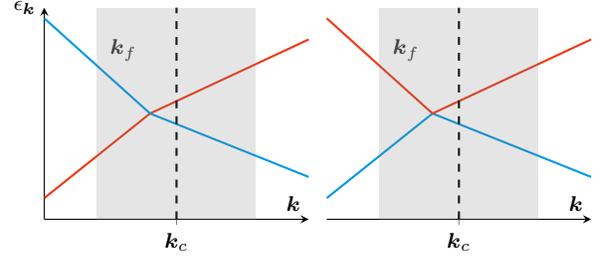
\begin{figure}
    \begin{tikzpicture}
    \begin{axis}
        [
            width=0.6\columnwidth,
            axis lines = center,
            xtick = {2.5},
            xticklabels={$\bvec k_c$},
            ytick = \empty,
            xmin = 0, xmax = 5,
            ymin = 0, ymax=2,
            xlabel = $\bvec k$, ylabel = $\epsilon_{\bvec k}$,
            ylabel style={left},
            clip = false
        ]
        \addplot[thick, mark=none, c-channel] coordinates {(0,0.2) (2,1) (5,1.7)};
        \addplot[thick, mark=none, p-channel] coordinates {(0,1.9) (2,1) (5,0.4)};
        \addplot[thick, mark=none, dashed] coordinates {(2.5, 0) (2.5, 2)};
        \addplot[fill=gray, mark=none, fill opacity=0.2, draw=none]
        coordinates {(1,0) (1,2) (4,2) (4,0)};
        \node at (axis cs:1.5,1.6) {\textcolor{darkgray}{$\bvec k_f$}};
    \end{axis}%
    \end{tikzpicture}~
    \begin{tikzpicture}
    \begin{axis}
        [
            width=0.6\columnwidth,
            axis lines = center,
            axis y line = none,
            xtick = {2.5},
            xticklabels={$\bvec k_c$},
            ytick = \empty,
            xmin = 0, xmax = 5,
            ymin = 0, ymax=2,
            xlabel = $\bvec k$, ylabel = $\epsilon_{\bvec k}$,
            ylabel style={left},
            clip = false
        ]
        \addplot[thick, mark=none, p-channel] coordinates {(0,0.2) (2,1) (5,0.4)};
        \addplot[thick, mark=none, c-channel] coordinates {(0,1.9) (2,1) (5,1.7)};
        \addplot[thick, mark=none, dashed] coordinates {(2.5, 0) (2.5, 2)};
        \addplot[fill=gray, mark=none, fill opacity=0.2, draw=none]
        coordinates {(1,0) (1,2) (4,2) (4,0)};
        \node at (axis cs:1.5,1.6) {\textcolor{darkgray}{$\bvec k_f$}};
    \end{axis}%
    \end{tikzpicture}
    \caption{\captiontitle{Problem with band crossings and Refinement.}
    When refining in band space representation the band structure shown in the
    left plot we ignore the fact that a band crossing occurs within the region
    around the coarse mesh point.
    The orbital-to-band transformations, calculated only at the points of the
    coarse mesh point will map the averaged band into the orbital space
    respective their order at the coarse point.
    }
    \label{fig:A_band_crossing_problem}
\end{figure}

To illustrate the issue of evaluating the average in band space and transforming
into orbital space afterwards we imagine a region of the BZ where multiple bands
cross (cf.~\cref{fig:A_band_crossing_problem}). Mathematically, we can formulate
the problem that arises when the mapping from orbitals to bands changes as
\begin{align}
    \begin{split}
        & \sum_{\bvec k_f} \sum_{b_1b_2}
            u^{*o_1b_1}(\bvec l+\bvec k_f)
            u^{*o_3b_2}(\bvec l'\pm \bvec k_f) \\
            & \qquad \dot L_{\pm}^{\Lambda,b_1b_2}(\bvec l + \bvec k_f, \bvec k - \bvec k_f)
            u^{o_3b_1}(\bvec l + \bvec k_f)
            u^{o_4b_2}(\bvec l \pm \bvec k_f) \\
        & \neq \sum_{b_1b_2}
        u^{*o_1b_1}(\bvec l)
        u^{*o_2b_2}(\bvec l')
        u^{o_3b_1}(\bvec l)
        u^{o_4b_2}(\bvec l') \\
        &\qquad \sum_{\bvec k_f}
        \dot L^{\Lambda,b_1b_2}_{\pm}(\bvec l+\bvec k_f, \bvec l' \pm \bvec k_f )
    \end{split}
\end{align}

\subsubsection*{Summed Symmetries:}
Applying a symmetry to the Hamiltonian reads
\begin{equation}
    S_{\bvec k} H(\bvec k ) = U^\dagger (\bvec k) H(\bvec k) U(\bvec k) \,,
\end{equation}
where $U$ provides the needed phase-shifts and orbital maps for the
transformation of orbitals across BZ-boundaries. From this we can deduce the
following relation:
\begin{align}
    \begin{split}
        S_{\bvec k}&\Bigg[\sum_{\bvec k_f} H(\bvec k\pm \bvec k_f)\Bigg] \\
         &= U^\dagger(\bvec k) \Bigg(
           \sum_{\bvec k_f} H(\bvec k \pm \bvec k_f) \Bigg)
         U(\bvec k) \\
        &\neq \sum_{\bvec k_f} U^\dagger(\bvec k \pm \bvec k_f) H(\bvec k \pm \bvec k_f) U(\bvec k \pm
        \bvec k_f)\\
        &= \sum_{\bvec k_f} S_{\bvec k \pm \bvec k_f} \left[ H(\bvec k \pm \bvec k_f) \right] \\
        &= \sum_{\bvec k_f} \left[ H(\bvec k \pm \bvec k_f) \right] \,.
    \end{split}
\end{align}
Having summed over the phases of the refined points we are unable to
identify the appropriate transformation matrices.
This problem does not arise if the averaging is performed in band space as the
phases introduced by the transformation matrices are zero -- the Hamiltonian in
band space corresponds to the energies.

One possible solution for this problem is thus averaging in band space
(averaging the energy) and then applying the orbital-to-band transformation of
the coarse point. Fortunately there exists an orbital-space solution to properly
correct for these quantitative effects in the loop derivative, presented below.

\subsection*{Re-symmetrizing the refined loops from orbital space}

Given a multi-site tight-binding system described by hopping amplitudes as a
function of site indices, the matrix $U_S(\bvec k)$ can analytically be
constructed for each point-group symmetry $S$ of the system. Note that the
real space unit cell \emph{must} be chosen such that its origin aligns with the
system's rotational symmetries. Otherwise, rotations (and other point-group
symmetries) are only symmetries of the system with additional real space
displacements (i.e. momentum-space complex phase shifts). The procedure to
obtain the matrix $U_S(\bvec k)$ is straightforward and shortly presented in the
following:
\begin{enumerate}
  \item For each site index $o$ of the tight-binding Hamiltonian apply the
    symmetry $S$ to its real space position $\bvec r_o$ by calculating
    $\tilde{\bvec r}_o = \hat S\bvec r_o$.
  \item Find the site index $o'$ and integer vector $\vec \imath$ (of dimension
    `dimensionality' $D$) such that $\bvec r_{o'} = \vec \imath \cdot \vec{\bvec
    R} + \tilde{\bvec r}_o$ with $\vec{\bvec R} = (\bvec R_1, \dots, \bvec
    R_D)^\mathrm{T}$ the system's lattice vectors.
  \item Save above information in a map $\bvec X_S(o)$ that takes the site
    index $o$ and returns the vector $\bvec X_S(o)$ with $\bvec X_S^1(o) = o'$,
    $\bvec X_S^{2,\dots,D+1}(o) = \vec\imath$.
  \item Set the elements of the transformation matrix to $U_S^{o,\bvec
    X^1(o)}(\bvec k) = \exp\bigg[i\hat S\bvec k\cdot\big(\sum_{l=1}^D\bvec
    X^{1+l}(o)\bvec R_l\big)\bigg]$.
\end{enumerate}
As (free) Green's functions transform in the same way as the tight-binding
Hamiltonian under symmetries, we have 
\begin{align}
  S\big[G(\bvec k)\big] = U^\dagger_S(\bvec k) G(\hat S(\bvec k)) U_S(\bvec k).
\end{align}
From this, we can follow that the loops must transform as two Green's functions
in their respective orbital and momentum indices. For both the particle-particle
and particle-hole loops, this amounts to
\begin{multline}
  S\big[L^{o_1o_2o_3o_4}_{\pm}(\bvec k_1,\bvec k_2)\big] = \sum_{o_1'o_2'o_3'o_4'}
  U_S^{*,o_1o_1'}(\bvec k_1) U_S^{o_3o_3'}(\bvec k_1) \\
  U_S^{*,o_2o_2'}(\bvec k_2) U_S^{o_4o_4'}(\bvec k_2)
  L^{o_1'o_2'o_3'o_4'}_{\pm}(\hat S \bvec k_1,\hat S \bvec k_2) \,.
\end{multline}
Above equation holds if the site indices $o_1$ and $o_3$ correspond to one
Green's function and $o_2$ and $o_4$ to the other, with $o_1,o_2$ being ingoing
and $o_3,o_4$ outgoing legs.

With above transformation rule at hand, we can derive a straightforward formula
used for re-symmetrizing the refined loop. Let the system's symmetry be
described by a point group $\mathcal G$ of order $N_{\mathcal G}$. The
re-symmetrized refined loop $L^{\mathcal{S}}$ follows as
\begin{align}
  L^{\mathcal{S},o_1o_2o_3o_4}_{\pm}(\bvec k_1,\bvec k_2) =
  \frac1{N_{\mathcal G}} \sum_{S\in \mathcal G} S
  \big[L^{o_1o_2o_3o_4}_{\pm}(\bvec k_1,\bvec k_2)\big] \,.
\end{align}

\section{Multi-site TUFRG and symmetries}
\label{sec:A_tu_multisite}
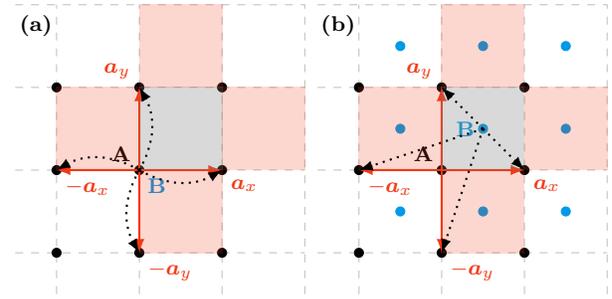
\begin{figure}[ht]
    \begin{tikzpicture}[scale=0.55]
    \coordinate (Origin)   at (0,0);
    \pgftransformcm{1}{0}{0}{1}{\pgfpoint{0cm}{0cm}}
    \coordinate (Bone) at (0,2);
    \coordinate (Btwo) at (2,0);
    \draw[style=help lines,dashed, lightgray] (-3,-3) grid[step=2cm] (4,4);

    \foreach \x in {-2,-0,...,2}{
      \foreach \y in {-2,-0,...,2}{
        \node[draw, circle, inner sep=1.2pt, fill] at (\x,\y) {};
      }
    }
    \node[above left] at (0,0) {\textbf{A}};
    \node[below right, text=p-channel] at (0,0) {\textbf{B}};
    \draw [thick,-latex, c-channel] (Origin) -- (Bone)
        node [above left] {$\bvec a_y$};
    \draw [thick,-latex, c-channel] (Origin) -- (Btwo)
        node [below right] {$\bvec  a_x$};
    \draw [thick,-latex, c-channel] (Origin) -- (-2,0)
        node [below right] {$-\bvec a_x$};
    \draw [thick,-latex, c-channel] (Origin) -- (0,-2)
        node [below right] {$-\bvec a_y$};
    \filldraw[fill=gray, fill opacity=0.3, draw=none] (Origin) rectangle (2,2);
    \filldraw[fill=c-channel, fill opacity=0.2, draw=none] (Origin) rectangle (2,-2);
    \filldraw[fill=c-channel, fill opacity=0.2, draw=none] (2,0) rectangle (4, 2);
    \filldraw[fill=c-channel, fill opacity=0.2, draw=none] (0,2) rectangle (2, 4);
    \filldraw[fill=c-channel, fill opacity=0.2, draw=none] (-2,0) rectangle (0, 2);

    \draw [-latex, black, dotted, thick] (0,0) to [out=150, in= 30] (-2, 0);
    \draw [-latex, black, dotted, thick] (0,0) to [out=240, in=120] ( 0,-2);
    \draw [-latex, black, dotted, thick] (0,0) to [out= 60, in=300] ( 0, 2);
    \draw [-latex, black, dotted, thick] (0,0) to [out=330, in=210] ( 2, 0);
      \node[] at (-2.5,3.5) {\textbf{(a)}};
  \end{tikzpicture}
    \begin{tikzpicture}[scale=0.55]
    \coordinate (Origin)   at (0,0);
    \pgftransformcm{1}{0}{0}{1}{\pgfpoint{0cm}{0cm}}
    \coordinate (Bone) at (0,2);
    \coordinate (Btwo) at (2,0);
    \draw[style=help lines,dashed, lightgray] (-3,-3) grid[step=2cm] (4,4);

    \foreach \x in {-2,-0,...,2}{
      \foreach \y in {-2,-0,...,2}{
        \node[draw, circle, inner sep=1.2pt, fill] at (\x,\y) {};
        \node[draw=p-channel, circle, inner sep=1.2pt, fill=p-channel] at (\x+1,\y+1) {};
      }
    }
    \node[above left] at (0,0) {\textbf{A}};
    \node[left, text=p-channel] at (1,1) {\textbf{B}};
    \draw [thick,-latex, c-channel] (Origin) -- (Bone)
        node [above left] {$\bvec a_y$};
    \draw [thick,-latex, c-channel] (Origin) -- (Btwo)
        node [below right] {$\bvec a_x$};
    \draw [thick,-latex, c-channel] (Origin) -- (-2,0)
        node [below right] {$-\bvec a_x $ };
    \draw [thick,-latex, c-channel] (Origin) -- (0,-2)
        node [below right] {$-\bvec a_y $ };

    \filldraw[fill=gray, fill opacity=0.3, draw=none] (Origin) rectangle (2,2);
    \filldraw[fill=c-channel, fill opacity=0.2, draw=none] (Origin) rectangle (2,-2);
    \filldraw[fill=c-channel, fill opacity=0.2, draw=none] (2,0) rectangle (4, 2);
    \filldraw[fill=c-channel, fill opacity=0.2, draw=none] (0,2) rectangle (2, 4);
    \filldraw[fill=c-channel, fill opacity=0.2, draw=none] (-2,0) rectangle (0, 2);

    \draw [-latex, black, dotted, thick] (1,1) -- (0,-2);
    \draw [-latex, black, dotted, thick] (1,1) to (2, 0);
    \draw [-latex, black, dotted, thick] (1,1) to (0, 2);
    \draw [-latex, black, dotted, thick] (1,1) -- (-2,0);
      \node[] at (-2.5,3.5) {\textbf{(b)}};
  \end{tikzpicture} \\
  \caption{
      \captiontitle{Example of symmetry breaking in TUFRG.} Here we want to show
      the symmetry breaking introduced in TUFRG when analyzing a lattice with
      basis. The red boxes represent the unit cells of the lattice. We consider
      the nearest-neighbor interaction from a B to an A site as an example.
      \captionlabel{a} Square lattice with basis in the proper gauge
      representation. The formfactor expansion of the nearest-neighbor onsite
      interaction will consist of the indicated couplings. \captionlabel{b}
      Physical representation of the couplings indicated in \captionlabel{a}. It
      is apparent that this mixes different length scales and no longer respects
      the symmetry of the lattice.}
  \label{fig:sublattice_structure_tu}
\end{figure}

When using the TUFRG approximation for multi-site systems we need to be aware
that we introduce a mixing of differing length scales in the interaction. This is
exemplified in \cref{fig:sublattice_structure_tu} where we can clearly see that
the proper-gauge representation of the interaction (all sites
Fourier-transformed with respect to the same position) allows the definition of
the formfactor basis only with respect to the remaining A positions. The
realspace representation of the first shell is drawn in
\cref{fig:sublattice_structure_tu} \textbf{(a)}, where we also indicate an
exemplary interaction from a B-site to all neighboring A-sites.
If we now transform into the physical picture in \textbf{(b)} we can clearly see
that the seemingly consistent interaction length mixes differing length scales.
If we transformed into the formfactor basis the different length scales would have
been integrated into the same formfactor shell, loosing the ability to
differentiate their contributions.
In the cross-channel projections this introduces a slight symmetry
breaking into the approximation. We see breaking of otherwise degenerate
states as well as slight modifications of critical scales.

This behavior can be rectified by introducing a site-dependent formfactor
expansion. We need to define a different set of neighboring formfactors
for the A and B site in the above example.

\section{Formfactor generation}
\label{sec:A_formfactor_generation}
We want to detail the generalization of the formfactor generation method
described in Appendix 3 of Ref.~\onlinecite{platt-hanke-thomale-2013}.
In order to do this we first recall the methodology of that publication before
expanding to our new approach.

\subsection*{Original Method}
Given a symmetry Group $\mathcal G$ of the problem with group elements $g$ and
representations $\Gamma_i$ with characters $\chi_i(g)$
we define the projectors
\begin{equation}
  \mathcal P(\Gamma_i) = \sum_{g\in\mathcal G} \chi_i^*(g)g,
\end{equation} into the representation's contribution.

To now find the shell-basis for this representation apply the projector to a
trial bond taken from the shell $n$ of the realspace lattice $\phi_n(\bvec r) =
\delta_{i,i+\bvec r}$. The result is the realspace-representation of the
formfactor:
\begin{equation}
    \varphi^{(n)}_{i}(\bvec r)
        = P(\Gamma_i) \phi^{(n)}(\bvec r)
        = \sum_{g \in \mathcal G} \chi_i^*(g)g \, \phi^{(n)}_j(\bvec r) \,,
    \label{eqn:ff_proj}
\end{equation}
a simple Fourier transformation yields the desired momentum space formfactors:
\begin{align}
  \varphi_{f}(\bvec k) &{}= \sum_{\bvec r \in \mathrm{Bonds}}
        e^{-i\bvec k\cdot\bvec r}\varphi_{f}(\bvec r) \,,\\
    \varphi_{f}(\bvec r) &{}= \bzint k \, e^{i\bvec k\cdot\bvec
    r}\varphi_{f}(\bvec k)\,.
    \label{eqn:ff_fourier}
\end{align}
For the multi-dimensional representation we use a number of different trial
bonds equal to the dimensionality.
This is supposed to ensure that the number of formfactors is always sufficient
to provide a unitary transformation from a realspace lattice shell into 
formfactor space.
This however breaks down at longer ranges as will be seen in the next section.

\subsection*{Improved Version}
The issues with the above mentioned method arise from its non-generality, the
dimensionality of the representations is given but we encounter momentum shells
which have more points than we have representations in the symmetry group (the
fourth shell of the square lattice is an example). For these shells the above
mentioned procedure does not generate a sufficient number of formfactors for the
unitary transformation of the spaces.
For these we are able to find multiple symmetry-inequivalent formfactors for a
given representation. To produce these formfactors while maintaining
orthogonality between the formfactors we use the following procedure:

For \emph{all} bonds of a given length in the realspace lattice we apply the projector
defined above for \emph{all} representations.
We reduce the obtained (too large) set of realspace formfactors $\varphi (\bvec r)$
to a linearly independent subset and orthogonalize using the Gram
Schmidt orthogonalization procedure.

We can then promise the following equations for the formfactor basis:
\begin{align}
    \label{eqn:ff_unity}
  \delta(\bvec k - \bvec k') &{}
  = \sum_{f} \varphi_{f}(\bvec k) \varphi^*_{f}(\bvec k') \,,\\
  \delta_{f,f'} &{}= \bzint k \,\varphi_{f}(\bvec k)
    \varphi_{f'}(\bvec k)
    \,.
\end{align}
To ensure we generate the physical set of formfactors (the ones described in
Ref.~\onlinecite{platt-hanke-thomale-2013}) we use the following iterative algorithm for
the generation of formfactors, this favors the formfactors generated by the
original method.

\begin{algorithm}[H]
    \caption{Generate all formfactors of shell $n$}
    \begin{algorithmic}[1]
        \State $F(n) = \emptyset$
        \While{$N_{\varphi}(n)<N_{\mathrm{Bond}}(n)$}
        \For{$\bvec b \in n$}
        \For{$\Gamma_i \in \mathcal G$}
            \State $\varphi(\bvec r) = \mathcal P(\Gamma_i) \bvec b$
            \If{$\varphi(\bvec r) \neq 0$ \textbf{and}
                $a\varphi(\bvec r) \not\in F(n) \forall a\in\mathbb{C}$}
                \State $F = F \cap \mathcal F[\varphi(\bvec r)]$;
            \EndIf
        \EndFor
        \EndFor
    \EndWhile
    \end{algorithmic}
\end{algorithm}

The number of formfactors which need to be chosen is highly dependent upon the
problem under scrutiny. If the interactions are highly localized, a few
formfactors will suffice to capture the primary dependencies. A more detailed
analysis of the necessary number of formfactors should be performed for each
simulation.

\section{Scaling tests}
\label{sec:scaling_tests}
Beyond the equivalence checks and reproduction of known results, we further
checked the correct scaling behavior of our implementations against exact
solutions for small systems. Therefore a short exact diagonalization code for
a 1D Hubbard chain, a 1D Rashba chain and 2D graphene has been implemented. We
then calculated the occupation number $\rho_{i,j} = \langle
c_{i}^{\dagger}c_j\rangle$ using ED and a RS-TUFRG
implementation in the single frequency
approximation \cite{weidinger-bauer-vondelft-2017,reckling-honerkamp-2018}.
The error is expected to scale proportionally to $U^3$. To verify the
correct scaling behavior, roughly $60$ bosonic frequencies were necessary and
the accepted error of the adaptive integrator has been set to $10^{-6}$, which
allows us to verify the proportionality down to an absolute error of $\approx
10^{-6}$.

\begin{figure*}
  \includegraphics[width=\linewidth]{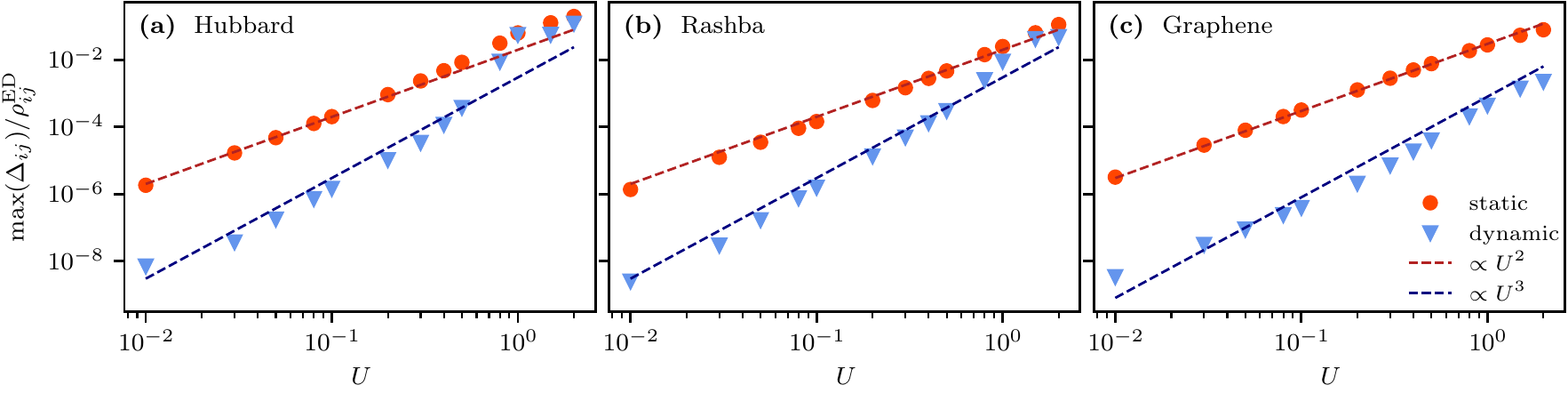}
  \caption{\captiontitle{Scaling error of the occupation matrix compared to exact
    diagonalization}
    We show the maximal error of the occupation number relative to the value of
    the exact diagonalization occupation number on the y-axis on a logarithmic
    scale.
    On the x-axis we show the interaction value on a logarithmic scale.
    As a guideline for the eye, we show the $U^2$ and $U^3$ error scalings
    expected from perturbative arguments.
    For the static calculations the error is expected to scale $\propto U^2$ and
    for the dynamic to scale $\propto U^3$.}
  \label{fig:GrapheneScaling_ED}
\end{figure*}

The results are summarized in \cref{fig:GrapheneScaling_ED}.
At low values of the interaction, the integration error leads to a deviation
from the expected scaling, as can be seen in all three cases. For larger U,
higher order terms can become important also deteriorating the scaling,
this can be seen especially in the two 1D systems. Apart from these minor
effects, the data follow the $U^3$-curve very closely.

\bibliographystyle{spphys}
{\footnotesize \bibliography{references}}

\end{document}